\documentclass[twocolumn]{aastex631}

\begin{document}

\title{Local two-dimensional simulations of the ignition of a helium shell detonation on a white dwarf by an impacting stream}

\author{Nethra Rajavel}
\affiliation{Department of Physics \& Astronomy, The University of Alabama, Tuscaloosa, AL, USA}

\author{Dean M. Townsley}
\affiliation{Department of Physics \& Astronomy, The University of Alabama, Tuscaloosa, AL, USA}

\author{Ken J. Shen}
\affiliation{Department of Astronomy and Theoretical Astrophysics Center, University of California, Berkeley, CA, USA}



\begin{abstract}

The double detonation model is one of the prevalent explosion mechanisms of Type Ia Supernovae (SNe Ia) wherein an outer helium shell detonation triggers a core detonation in the white dwarf (WD). The dynamically driven double degenerate double detonation (D$^6$) is the double detonation of the more massive WD in a binary WD system where the localized impact of the mass transfer stream from the companion sets off the initial helium shell detonation. To have high numerical resolution and control over the stream parameters, we have implemented a study of the local interaction of the stream with the WD surface in 2D. In cases with lower base density of the shell, the stream’s impact can cause surface detonation soon after first impact. With higher base densities, after the stream hits the surface, hot material flows around the star and interacts with the incoming stream to produce a denser and narrower impact. Our results therefore show that (1) a directly impacting stream for both a relatively high resolution and for a range of stream parameters can produce a surface detonation, (2) thinner helium shells ignite more promptly via impact, doing so sooner, and (3) there are lower limits on ignition in both shell density and incoming stream speed with lower limits on density being well below those shown by other work to be required for normal appearing SN Ia. This supports stream ignition and therefore the D$^6$ scenario, as a viable mechanism for normal SNe Ia.

\end{abstract}



\section{Introduction} \label{sec:intro}

Type Ia Supernovae (hereafter SNe Ia) are thermonuclear explosions of white dwarfs (WDs). Although it is known that these WDs disrupt, the exact mechanism by which this occurs remains uncertain. WDs near the Chandrasekhar mass limit (1.4 $\textup{M}_\odot$), known as Chandrasekhar mass WDs can explode as they gain mass and reach the limit but WDs less massive than this limit or sub-Chandrasekhar Mass WDs (sub-Ch WDs), can also form SNe Ia through a completely different explosion mechanism \citep[e.g.][]{Maoz}. 

The explosion mechanism of sub-Ch WDs is still a topic of debate. However, one of the prevalent models, the double detonation model, shows promise in being able to replicate the spectra of observed normal SN Ia \citep{Kromeretal,Townsley_2019,Boos_2021, Shen2021}. According to the double detonation model, there is an initial detonation in the outer thin helium shell of a carbon-oxygen (CO) WD which then ignites a detonation in the CO core \citep{WoosleyWeaver}. There has been a lot of work in the past on simulating the entire double detonation scenario \citep{Guillochonetal2010,Dan2011,Raskin2012,MollWoosley2013, Pakmor2013, Fenn2016, Garcia-Senz_2018, Townsley_2019, Boos_2021, Gronow2020, Gronowetal2021,Collins2022, Pakmor2022, Roy2022}. However, the helium detonation in many studies is artificially created using a hot spot \citep{Dan2015,Garcia-Senz_2018,Townsley_2019,Tanikawa2019}. The conditions necessary for helium shell ignition have been studied in the past \citep{Holcombe2013, Moore_2013} but the ignition mechanism is significantly less explored and modeled. 
\hfill

\begin{figure*}[ht!]
\centering
\gridline{\fig{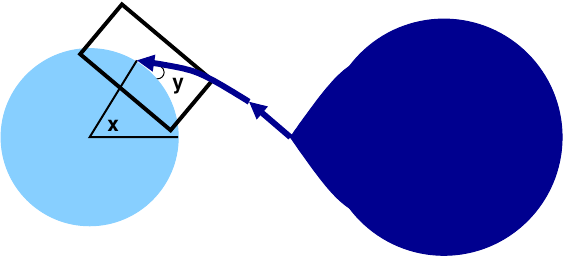}{0.44\textwidth}{(a)}
          \fig{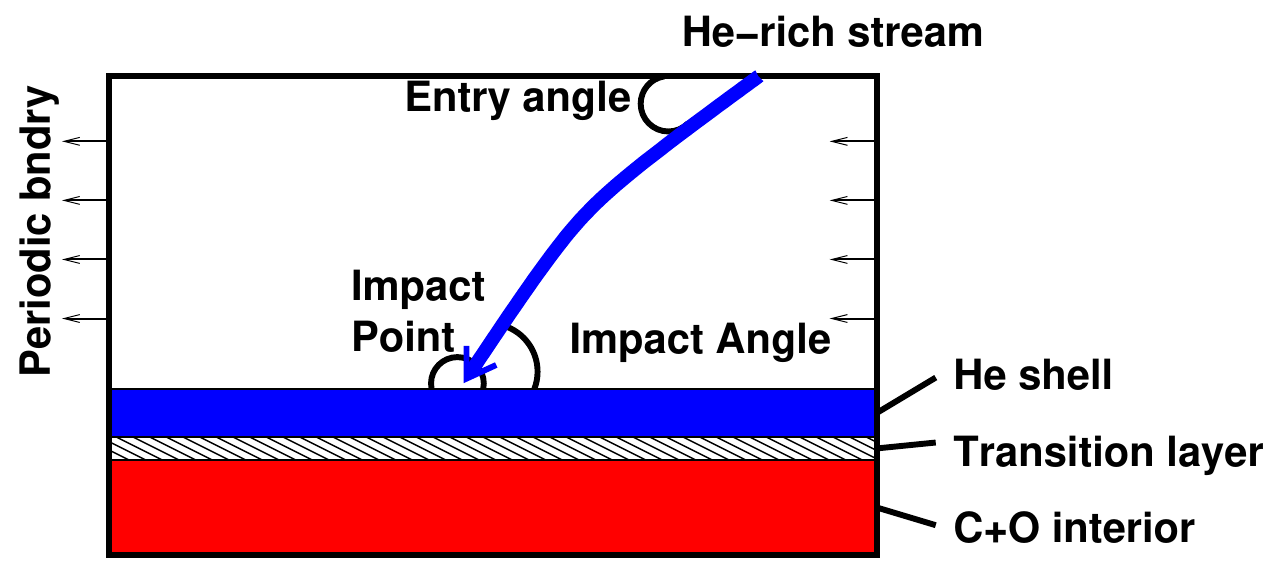}{0.45\textwidth}{(b)}}
\caption{(a) Dark blue region indicates the companion WD while the light blue region indicates the primary WD. Mass transfer from the companion results in a stream directly impacting the surface of the primary WD. Our simulation focuses on the stream striking the WD surface, a very local perspective to the surface of the primary WD. (b) The primary WD is represented as a flat plane with two layers: the outermost helium shell and the innermost carbon-oxygen core. The entire region represented here is the region within the black rectangle in (a). The angle marked as x is the impact location angle and the angle marked y is the impact angle.}
\label{fig:example}
\end{figure*}


One scenario for ignition of the helium shell is the dynamically driven double degenerate double detonation (D$^6$) scenario \citep{Guillochonetal2010,Dan2011,Pakmor2013,Shen2018b}. As the name suggests, the dynamical mass transfer onto the primary WD from the companion in a double degenerate system (a binary WD system) causes a double detonation in the primary WD. From here on, the primary WD refers to the higher mass WD and the companion WD refers to the lower mass WD. The companion WD being lower in mass and having a larger radius leads to it filling its Roche-lobe first thereby initiating mass transfer to the primary WD. The mass ratio of the primary and companion WD dictates if the mass transfer leads to the formation of an accretion disk or if the material from the mass transfer directly impacts the surface of the primary \citep{Lubow1975,Nelemans2001,Marsh_2004}. Specifically, the case where both the WDs in the binary system are CO WDs (0.45 - 1.0 \(\textup{M}_\odot\)) would result in the accretion stream directly impacting the surface of the primary WD and is thought to produce the outer helium shell detonation necessary for the D$^6$ model. 

\citet{Guillochonetal2010} simulated the mass transfer directly impacting the surface in a two star system in 3D. They found that a ``knot'' of dense hot material formed on the surface of the primary WD which eventually causes a surface detonation. The low resolution of the study was ultimately unable to investigate and characterize the ignition mechanics in detail. \citet{Pakmor2013} simulated the merger of two CO WDs where the accretion of material on the primary detonates the thin helium shell on it. However, the simulation was in a two-star system and hence the resolution was again lower than desired for a clear characterization of the local mechanism of the ignition.
\citet{Glasner2018} simulated the accretion of material from the companion and explored the possibility of a detonation in both thick (0.1 $\textup{M}_\odot$) and moderate (0.05 $\textup{M}_\odot$) shell cases on the primary WD (1.0 $\textup{M}_\odot$ WD). They observed a helium detonation for thick and moderate shell cases, and in the thick shell case they observed a transition to the core (edge-lit scenario). 
\citet{Pakmor2021} observed a unique result on simulating the merger of a ``hybrid'' HeCO WD (having an unusually thick helium shell) with a CO WD. The helium shell detonated via stream ignition but did not lead to a core ignition and instead travelled upstream to double detonate the companion. 
\citet{Iwata2022} studied the spontaneous helium shell ignition in 1D and found ignition challenging, highlighting the need for multi-dimensional investigations.  
\citet{Wong2023} simulated the stable mass transfer from the companion to the primary WD in 1D and were successful in producing a dynamical helium flash on the primary WD, an alternative to an impact ignition scenario. 

Most of the previous work listed either produced a surface ignition within a simulation including the whole binary system but with a low resolution or studied the accretion and ignition in 1D which doesn't address the stream impact. In this work, we will demonstrate a two-dimensional simulation of the direct impact of the accretion stream on the surface of the primary WD. Unlike previous work, this study is local to the surface of the primary WD to focus on the stream and the ignition mechanism. 
 
Section \ref{sec:method} is dedicated to the methods used to perform the study. Results found by varying stream characteristics are presented in section \ref{sec:results}. Section \ref{sec:heshell} discusses the outcome when the base density of the helium shell is varied. We summarize and conclude the paper in section \ref{sec:summary}.

\begin{table*}[ht!]
\begin{center}
\caption{Results from the two body + test mass trajectory integrator. $M_1$ and $R_1$ represent the mass and radius of the primary WD. Similarly, $M_2$ and $R_2$ represent the mass and radius of the companion WD.}
\begin{tabular}{|c|c|c|c|c|c|c|}
\hline
\multicolumn{1}{|l|}{$M_1$ (\(\textup{M}_\odot\))} & \multicolumn{1}{l|}{$M_2$ (\(\textup{M}_\odot\))} & \multicolumn{1}{l|}{$R_1$ ($ 10^8$ cm)} & \multicolumn{1}{l|}{$R_2$ ($10^8$ cm)} & \multicolumn{1}{l|}{Impact velocity ($10^8$\ cm\ s$^{-1}$)} & \multicolumn{1}{l|}{Impact angle ($^\circ$)} & \multicolumn{1}{l|}{Impact location angle ($^\circ$)} \\ \hline
1.0                                      & 0.45                 & 5                                       & 10                                          & 5.8                           & 45                             & 49 \\ \hline
1.0                                      & 0.6                  & 5                                       & 8.5                                        & 5.3                           & 52                             & 34 \\ \hline
1.0                                      & 0.7                  & 5                                        & 8                                          & 5.0                           & 55                             & 28 \\ \hline
1.0                                      & 0.8                  & 5                                       & 7.5                                        & 4.7                           & 57                             & 23 \\ \hline
\end{tabular}
\label{table:1}
\end{center}
\end{table*}

\section{Methods} \label{sec:method}

Our work is motivated by a double degenerate system with two CO WDs where the primary is more massive than the other. From \citet{Marsh_2004}, we know that the mass ratio of a system defines whether or not the accretion is direct impact or disk accretion. For a primary WD which is a more massive CO WD ($\approx 0.8 \textup{M}_\odot$ to $1.0 \textup{M}_\odot$), a companion CO WD would result in direct impact accretion. The direct impact would result in a stream, originating at the $L_1$ point, that strikes the surface of the primary WD. Both the WDs also have a helium outer shell; therefore the transferred mass or stream would mainly consist of helium.
 
We choose to focus on the interaction of the stream with the primary WD in order to follow the impact and effect of the stream and thereby the ignition mechanism. The model is simulated in two dimensions using Flash \citep{Fryxell}, a multiphysics reactive hydrodynamic simulation code. Two different nuclear reaction networks are used, a small 13-isotope nuclear reaction network \citep{Fryxell} and a large 55-isotope nuclear reaction network. The smaller reaction network, comprising $^{4}$He, $^{12}$C, $^{16}$O, $^{20}$Ne, $^{24}$Mg, $^{28}$Si, $^{32}$S, $^{36}$Ar, $^{40}$Ca, $^{44}$Ti, $^{48}$Cr, $^{52}$Fe, $^{56}$Ni, is used to conduct a preliminary investigation on the possibility of producing a surface detonation without utilizing a large computation time. Later, a larger 55-isotope reaction network, which was previously used in double detonation simulations \citep{Townsley_2019,Boos_2021,Gronowetal2021}, is utilized since this is required for ignition of the thinnest shells. Modules for Experiments in Stellar Astrophysics (MESA) is used in combination with Flash for the larger reaction network. 

Heat conduction is not included in our models. This is because our models use a helium shell with a base density of around $10^5$\ g\ cm$^{-3}$  and from \citet{Timmes2000} for that density, the deflagration speed would be $<$ 100\ cm\ s$^{-1}$ which would be too small to be noticeable on the scale of our simulation (approx 10s simulation time and domain length of $20 \times 10^8$\ cm) which is designed to observe the higher speeds attained by detonations.

There are two main components to our model, the CO WD surface with a helium shell and the impacting stream. We model the surface in a plane parallel approximation with the bottom layer representing the core and the top layer representing the helium shell of the WD as seen in Figure \ref{fig:example}.
We defer the curved surface simulation to future work.
The computational configuration for the WD surface is similar to that of \citet{Moore_2013} and uses a surface gravity of $4 \times 10^{8}$\ cm\ s$^{-2}$ and a density of $5 \times 10^5$\ g\ cm$^{-3}$ at the base of the helium layer (this is varied later in Section \ref{sec:heshell}). This surface gravity approximately corresponds to a 1.0 $\textup{M}_\odot$ WD. The scale height, H, is approximately 700\ km. The base of the helium layer is taken to be $1.5 \times 10^8$\ cm from the bottom of the computational domain. 

The boundary condition on the bottom of the grid is a variation of a hydrostatic boundary condition \citep{Zingale2002} where the normal velocity is handled by a reflective boundary condition. We have not observed any issues with handling the normal velocity in such a manner but we will make modifications if they are needed in future simulations. The other boundary conditions will be discussed in the subsections below. The length of the surface in the horizontal direction mimics the circumference of a 1.0 $\textup{M}_\odot$ WD. With a radius of $\approx 5 \times 10^8$\ cm, the circumference would be $\approx 30 \times 10^8$\ cm. A surface length of $20 \times 10^8$\ cm was initially used to reduce computational time and, later, the larger surface length of $30 \times 10^8$\ cm was used. A greater surface length is only expected to delay the ignition and hence is not a defining factor. Adaptive mesh refinement is used, with refinement triggered on variations in density, temperature, and helium mass fraction.  The largest cell sizes are square and $\approx 310$\ km in each dimension, and refinement is allowed to decrease the cell size (by splitting cells in half in each dimension) to a minimum cell size of $\approx 20$\ km.


The stream, formed by the Roche-lobe overflowing companion, strikes the surface of the primary WD as pictured in Figure 1(a) with a velocity and impact angle dependent on the binary system. These are parameters that are necessary to define the stream dynamics in our model but are also parameters that are specific to a given binary system. Here, we refer to the angle between the stream and a line tangent to the surface as the impact angle (can be seen in Figure \ref{fig:example} (a)). In order to obtain sensible estimates of velocity and impact angle, we generated a simple test mass trajectory integrator which models the evolution of a test mass in a binary system in the rotating frame by considering the gravitational force on the test mass, the Coriolis force and the centrifugal force. Two bodies (WDs) are placed such that the separation distance between them is equivalent to the distance between the WDs when the companion WD would fill its Roche-lobe. We equate the radius of the companion WD to the radius of the Roche-lobe of the companion star and find the separation distance using the equation for the radius of the Roche-lobe of the second star from \citet{1967AcA....17..287P}. The orbital period is then determined from the chosen masses and the resulting separation.

The mass-less test particle is placed slightly away from the first Lagrange point ($L_{1}$) of the system such that it is on the side of the primary WD. The test particle represents the stream particles that would be transferred to the primary WD when the companion WD fills its Roche-lobe (which is why we position this particle near $L_{1}$). The motion of the test particle was simulated until it crosses the radius corresponding to the primary WD's surface. The primary WD is a 1.0 $\textup{M}_\odot$ WD and we vary the mass of the companion WD from 0.45 to 0.8 $\textup{M}_\odot$. The radii of the WDs are taken from \citet{Hamada1961}. The motion of the test particles varies for systems with different mass ratios as observed in Table \ref{table:1} which lists the results from this experiment. The impact velocity, impact angle of the test mass and the angle from $L_1$ are the main results from this experiment shown in the last three columns.  The angle subtended at the center of the primary by the point of impact and $L_1$ describes the position of the impact, referred to from now on as impact location angle. There is an increase in impact angle and decrease in velocity with the increase in the mass of the companion WD. Also, there is a decrease in the impact location angle as the mass of the companion WD increases.  

The material in the “stream” might not be uniform or have a consistent flow. Therefore, a reasonable possibility to consider would be that the stream is composed of smaller chunks of material. In fact \citet{Guillochonetal2010} reason that the stream would consist of ``knots'' which would have a size similar to the stream radius. For that reason, we also simulate the impact of a single mass of material on the WD surface and discern if it can detonate the surface before constructing a stream of material that strikes the surface. Additionally, even though Table \ref{table:1} gives us insight on the stream parameters, the single impact study is helpful to find stream parameters that detonate the helium shell. Hence our work can be broken down into two main sections, (1) single impact and (2) the stream impact.

 \subsection{Single Impact Study}

 As discussed previously, a single impact on the WD surface can assist in understanding the behavior of a stream impact and aid in deducing the stream parameters necessary to produce a surface detonation. To do this we use a circular mass of helium. The impact of a circular mass is technically equivalent to a cylindrical mass striking the surface instead of a spherical mass but performing the simulation in 2D limits the geometry of the impacting mass. 
 
 The density and temperature of this impacting mass was initially determined using the density and temperature of the background material that fills the space above the surface which are $\rho_{\rm{fluff}}$ and $T_{\rm{fluff}}$ respectively. In order for the size of the circular mass to not expand suddenly as soon as the simulation starts, it must be in approximate pressure equilibrium with the surrounding material. To do this the density was fixed at a value comparable to the outer layers of the surface ($2.5 \times 10^4$\ g\ cm$^{-3}$) and the temperature was scaled so that an ideal gas would be in equilibrium. To decrease the discontinuity at the edge of this circular mass, a density gradient defined by 
 
 \begin{equation}
    \rho = \rho_{\rm{mass}} - (\rho_{\rm{mass}} - \rho_{\rm{fluff}})\times \left(\frac{d_{\rm{mass}}}{r_{\rm{mass}}}\right)^2 ,
\end{equation}
was used. Here, $\rho_{\rm{mass}}$ is the central density and is $2.5 \times 10^4$\ g\ cm$^{-3}$, $\rho_{\rm{fluff}}$ is the density of the background ``fluff'' material and is $\approx 10^{1.4}$\ g\ cm$^{-3}$. The density decreases from $\rho_{\rm{mass}}$ at the center of the mass to $\rho_{\rm{fluff}}$ at the edges. $r_{\rm{mass}}$ is the radius of the mass and $d_{\rm{mass}}$ is defined as  $\sqrt{((x-x_1)^2+(y-y_1)^2 )}$ where $x_1$ and $y_1$ are the coordinates of the center of the circular mass. The temperature was scaled with respect to the density gradient and is defined as the following: \begin{equation}
    T = \frac{\rho_{\rm{fluff}} \times T_{\rm{fluff}}}{\rho}
\end{equation}
where T$_{\rm{fluff}}$ is the the temperature of the background material, $10^{6.4}$\ K. The base density of the helium shell and the surface gravity are not varied. A surface length of $20 \times 10^8$\ cm was used here to reduce computational time as mentioned previously. 

The radius, velocity and entry angle of the circular mass were varied to test whether helium shell detonates or not. The entry angle is defined as the angle at which the stream enters the domain as seen in Figure \ref{fig:example} (b). An entry angle of 90$^\circ$ would correspond to an impact that is perpendicular to the surface. Although the results of the impact angle from the trajectory integrator are used as guidelines for determining the entry angle, the entry angle and impact angle are not the same. This is because the direction of the stream changes as it curves toward the surface due to gravity.

To avoid unwanted interaction of the detonation with underlying layers which could lead to ignition of a carbon detonation, the carbon and oxygen in the core were substituted with magnesium for this study. The small 13-isotope reaction network was utilized because this is only a preliminary study. The shell and circular mass are made out of pure helium.

\begin{figure}[ht!]
\centering
\includegraphics[width=0.45\textwidth]{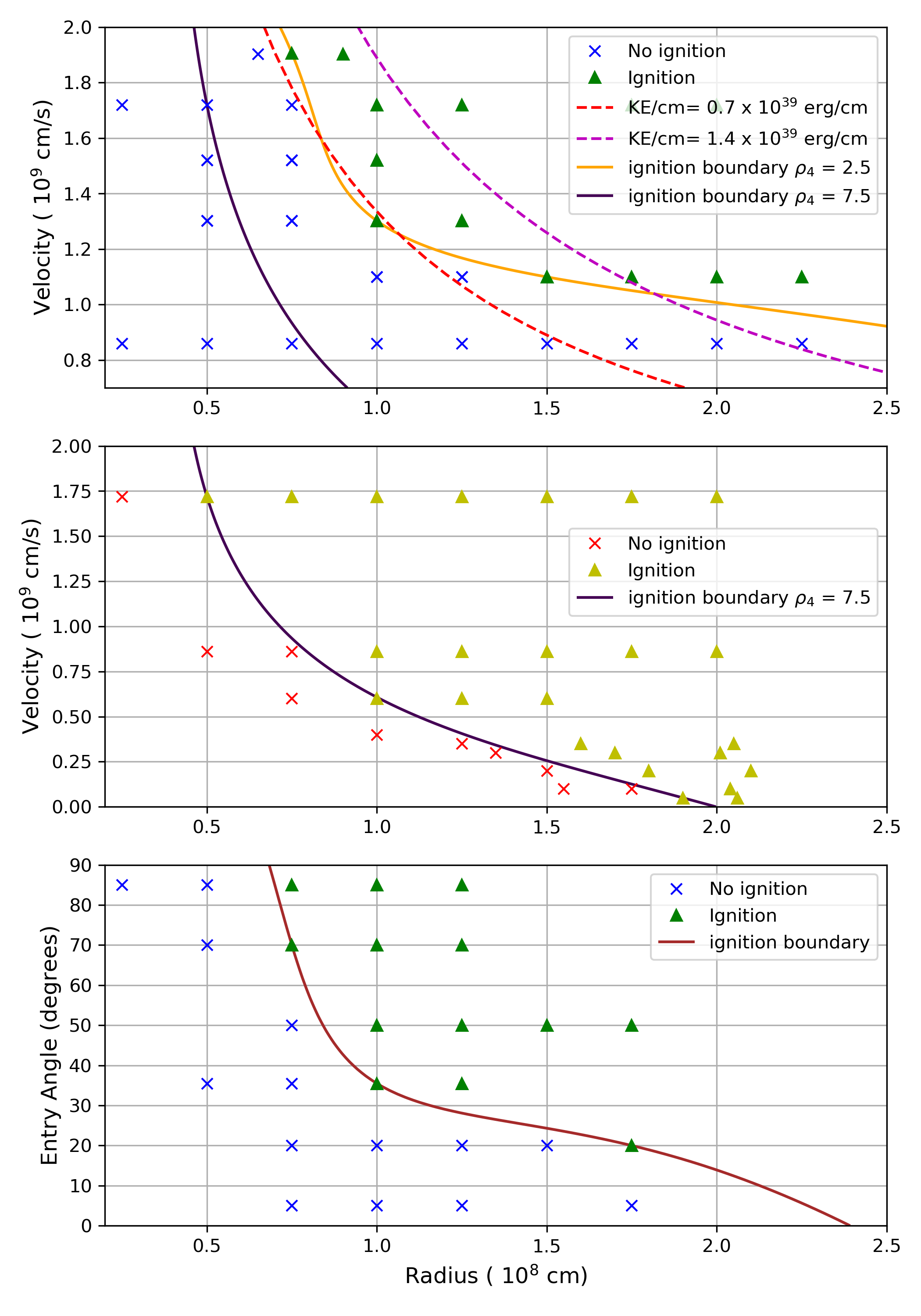}
\caption{Ignition boundary estimates for a circular mass hitting the WD surface. (Top) Ignition boundary estimates for $\rho_4$ = 2.5 when velocity is varied. (Middle) Ignition boundary estimates for $\rho_4$ = 7.5 when velocity is varied. (Bottom) Ignition boundary estimates for $\rho_4$ = 2.5 when entry angle is varied for a velocity of $1.3 \times 10^9$ ~cm~s$^{-1}$. }
\label{fig:plots}
\end{figure}

\subsection{Stream study}

In addition to the single impact study we proceeded to the more complicated stream study. The existing setup is preserved but, instead of the single impacting mass, a stream of material is designed to enter the grid by modifying the top boundary condition. A zero gradient boundary condition is defined everywhere on the top edge except within the radius of the stream where the material is given a specific density and velocity to impose an inflow onto the domain. The left and right edges are set to periodic boundary conditions to allow material to flow around the surface of the WD and mimic a spherical, continuous surface. Unlike the single impact study where a density gradient is defined to smooth the edges, we use a uniform density of $2.5 \times 10^4$\ g\ cm$^{-3}$ and a temperature of $2.5 \times 10^6$\ K for the stream. A density gradient is harder to implement in an inflowing stream but we will attempt this in future work. Our choice of stream parameters is influenced by the single impact study. 

The CO core of the WD is once again substituted with a magnesium core to avoid ignition of the underlying layer and the smaller, 13-isotope, nuclear reaction network is used in a preliminary stream study. The simulations are initially performed with the smaller surface length of $20 \times 10^8$\ cm and later with $30 \times 10^8$\ cm. We also try to observe the process in a higher resolution and therefore use a cell size as small as $\approx$ 5\ km after initial results where the smallest resolution is $\approx$ 20\ km.

After successfully obtaining a surface detonation with a small reaction network and magnesium core, we repeat the process for the more realistic case of a carbon-oxygen core and the larger 55-isotope reaction network. The abundance of the CO WD considered here has a similar abundance to the WD model in \citet{Townsley_2019}. The abundance of carbon and oxygen in the core are 0.4 and 0.6 respectively while the shell and stream have a nitrogen abundance of 0.009 and a helium abundance of 0.991. The proton capture on nitrogen is the slowest reaction in the CNO cycle. Therefore, small amounts of nitrogen are expected in the shell and stream, regions that have burned hydrogen to helium, through the CNO cycle. Nitrogen was added in the shell and stream for this reason and also because it has been shown that the presence of nitrogen in the shell can decrease the size of the minimum detonatable hotspot \citep{Shen_2014}. The length of the plane does not affect anything other than how long it takes for the helium shell to detonate. Since the larger reaction network requires a larger computation time per step and cell we therefore utilize a smaller plane length of $20 \times 10^8$\ cm.

\begin{figure*}[ht!]
\centering
\includegraphics[width=\textwidth]{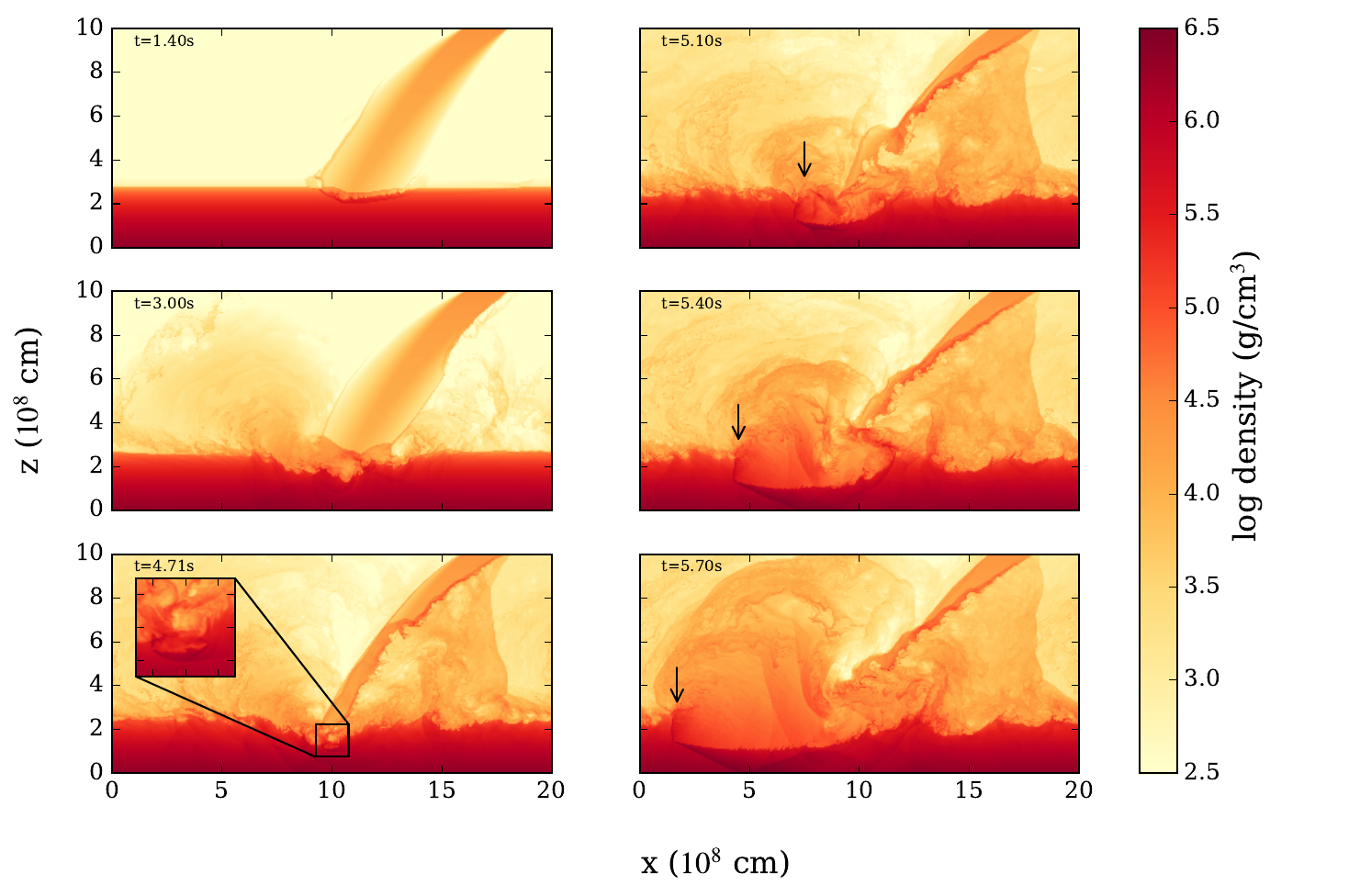}
\caption{Snapshots of density at various times, indicated on each panel, for a
two-dimensional stream (a flowing sheet) impacting the surface of a WD. The depicted case has a stream velocity = $6 \times 10^8$\ cm\ s$^{-1}$, a stream density = $2.5 \times 10^4$\ g\ cm$^{-3}$, a stream angle of 35.5$^\circ$, a stream half-width of $10^8$\ cm, $\rho_{base}$ of $5\times10^5$\ g\ cm$^{-3}$ and a cell size of $\approx 5$\ km, our highest resolution. At t=1.4\ s the stream impacts the WD surface and at $t=3.0$\ s the material that wraps around the surface, via the horizontal periodic boundary condition, begins to interact with the backside
of the stream, eventually constricting the stream to a narrower width. At t=4.71\ s the surface is ignited and next few time steps show the propagation of the detonation indicated by an arrow in each panel. An animated plot of this evolution is available from the Zenodo link at the end of the manuscript.}
\label{fig:dens_approx13}
\end{figure*}

\section{Results} \label{sec:results}

A surface detonation can be identified through several distinctive features. Firstly, the ignition region will reach temperatures as high as 2$\times 10^9$\ K, significantly exceeding the temperature of its surroundings. Over time, this high-temperature area will expand, bounded by the shock front. We use snapshots in density and temperature to detect a detonation in the helium shell. Although a density plot may not clearly indicate the higher density region during ignition, the propagation of the shock wave will be noticeable. When the temperature rises but fails to lead to a detonation, a hot spot may be observed, but it will dissipate after some time. In comparison, a self-propagating detonation wave will have a shock wave followed by a co-moving region of burning that supports it. We can compare fairly directly to the steady-state helium layer detonation structures computed by \citet{Townsley2012}. Both the single impact and the stream calculations demonstrated successful ignitions and show these features when a detonation occurs. The results of both these studies are presented below. 

\subsection{Single impact study} \label{subsec:blob}
 
Figure \ref{fig:plots} shows the collective results of the circular mass impact on the WD surface. The radius and entry velocity of the circular mass were varied and whether they detonate the surface or not is plotted in Figure \ref{fig:plots} (Top). The radius was varied between $2.5 \times 10^7$\ cm and $2.5 \times 10^8$\ cm. The velocity was varied between $8.6 \times 10^8$\ cm\ s$^{-1}$ and $1.91 \times 10^9$\ cm\ s$^{-1}$. The central density of the mass is fixed at $2.5 \times 10^4$\ g\ cm$^{-3}$. A small radius of less than $7.5 \times 10^7$\ cm does not detonate the surface for any entry velocity in this range. As the radius is increased, the velocity needed for a detonation decreases sharply at first and then decreases gently after a radius of $10^8$\ cm. A similar trend is observed when $\rho_{\rm{mass}}$ is increased to $7.5 \times 10^4$\ g\ cm$^{-3}$ (represented in Figure \ref{fig:plots} (Middle) and the solid dark purple curve) but a lower radius and entry velocity are observed as the ignition threshold. It is also interesting to observe the similarity in the constant kinetic energy per length curve and the boundary between the ignition and no ignition regions in Figure \ref{fig:plots} (Top). Since this is a 2D simulation, the density, area and the entry velocity of the mass were used to calculate the `kinetic energy per length' instead of the `kinetic energy'.  

Figure \ref{fig:plots} (Bottom) is the observed result when the entry angle and radius of the circular mass were varied but the entry velocity and $\rho_{\rm{mass}}$ were held constant at $1.3 \times 10^9$\ cm\ s$^{-1}$ and $2.5 \times 10^4$\ g\ cm$^{-3}$ respectively. The size threshold below which no entry angle is sufficient to allow ignition appears to be approximately $0.75\times 10^8$\ cm. The entry angle threshold for ignition decreases rapidly to about $30^\circ$ for impactor radii between 0.7 and $1.0\times 10^8$\ cm, then more slowly down to about $20^\circ$ for a radius of $2.0\times 10^8$\ cm.

The ignition boundaries shown correspond to decision boundaries created using a Support Vector Classifier (SVC) \citep{SVC} with a polynomial kernel of degree=3 and an independent coefficient `coef0'=3 and `coef0'=10 for a central density of $2.5 \times 10^4$\ g\ cm$^{-3}$ and $7.5 \times 10^4$\ g\ cm$^{-3}$ respectively. The decision boundaries suggests a transition between parameters that produce a surface detonation to those that do not. 

\begin{figure}[ht!]
\includegraphics[width=0.45\textwidth]{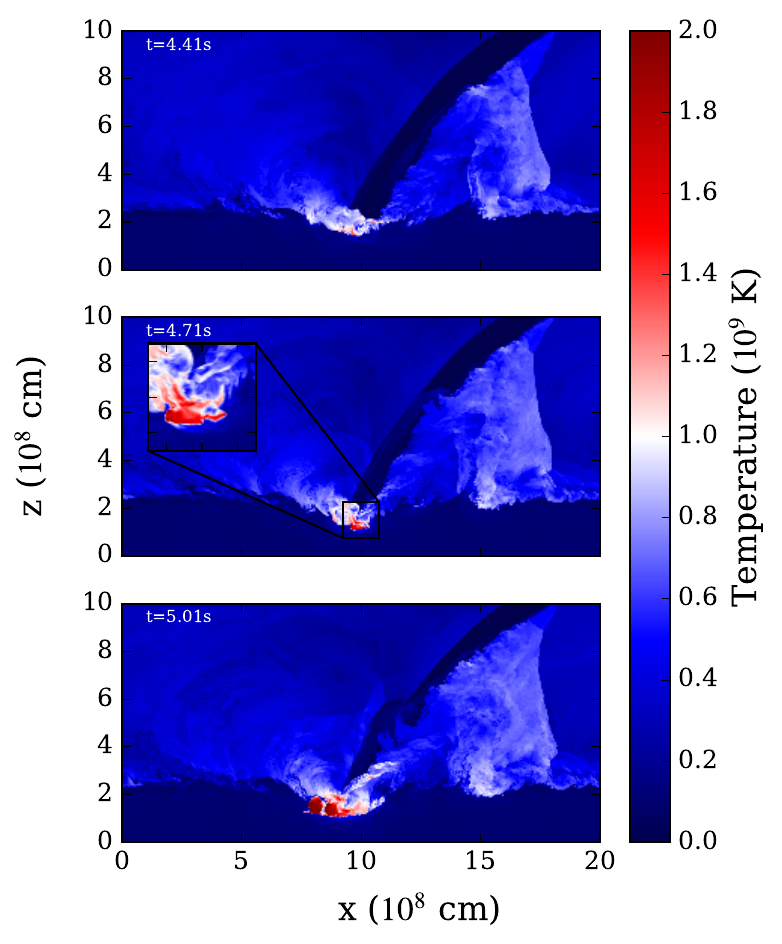}
\caption{Snapshots of temperature at times before, during and after and at ignition for a
two-dimensional stream impacting the surface of a WD. The case shown here has a stream velocity = $6 \times 10^8$\ cm\ s$^{-1}$, a stream density = $2.5 \times 10^4$\ g\ cm$^{-3}$, a stream angle of 35.5$^\circ$, a stream half-width of $10^8$\ cm, $\rho_{base}$ of $5\times10^5$\ g\ cm$^{-3}$ and a cell size of $\approx 5$\ km, our highest resolution. At t = 4.41\ s the stream is seen to be penetrating the surface as temperature rises. At t = 4.71\ s, a high temperature hotspot can be observed, indicating the ignition of the surface. At t = 5.01\ s the propagation of the detonation can be noticed. An animated plot of this evolution is available from the Zenodo link at the end of the manuscript.}
\label{fig:temp_approx13}
\end{figure}

\subsection{Stream study} \label{subsec:stream}

We choose a set of parameters for our stream based on the results of the single impact study. We choose parameters that are in the ignition region, but not far from the boundary and away from the extremes of the explored parameter space (Figure \ref{fig:plots}). The chosen parameters are a stream half-width of $10^8$\ cm, an entry velocity of $1.3 \times 10^9$\ cm\ s$^{-1}$, an entry angle of 35.54$^\circ$ and a stream density of $2.5 \times 10^4$\ g\ cm$^{-3}$. The base density of the helium shell is set to $5 \times 10^5$\ g\ cm$^{-3}$. A rough estimate of the rate of change of mass accumulated by the stream ($\dot{M}$) on the surface for these stream parameters is $\approx 0.0005$ $\textup{M}_\odot$\ s$^{-1}$  After a surface detonation was observed in the simulation for these initial values, the stream parameters were decreased to determine the lowest value at which a surface detonation is produced within 10 seconds runtime (run a little longer when no ignition is observed). We have simulated the stream impact with both a small reaction network with a magnesium core and pure helium shell and a large reaction network with a CO core and helium shell with nitrogen in it. These cases will be referred to as the ``pure helium'' case and the ``helium with nitrogen'' case from here on.

The formation and propagation of the surface detonation can be seen in Figure \ref{fig:dens_approx13}, which features snapshots in density for the pure helium case. Animated plots of several cases are available from the Zenodo link provided at the end of the manuscript. The phases of evolution are the same for the pure helium and helium with nitrogen cases, though some helium with nitrogen cases ignite earlier as will be discussed below. We demonstrate the pure helium case instead of the other because it is our highest resolution run (cell size $\approx$ 5\ km). Table \ref{table:2} lists all the different runs attempted by varying the velocity for the pure helium case and the helium with nitrogen case. It also lists the peak density at the point and time of ignition. The minimum $\tau_{nuc}/\tau_{dyn}$ just before ignition was found to be $\ll$ 1 for the pure helium shell case with a velocity of $6 \times 10^8$\ cm\ s$^{-1}$ (lowest resolution) [See appendix]. Therefore, the conditions at hot spots are feasible enough to induce ignition, without the influence of numeric ignition as emphasized by \citet{Glasner2018}.

Due to the two dimensional geometry, the inflowing material is physically in the form of a flowing sheet falling toward the WD surface. However, we will continue to use the term ``stream'' since we think it is more clear.
 
The stream flows onto the domain from the top right boundary and increases in width as it gets closer to the surface of the WD. Consequently, the density of the material initially impacting the surface is smaller than the stream density. The stream strikes the surface of the WD at about 1.4\ s after it enters the grid at t=0\ s. The material from the stream, which is pure helium in this case, bounces off the surface of the WD and continues to move away from stream. As more material strikes the surface, more material starts moving away from the stream, mostly in the leftward direction, and eventually wraps around the surface of the star, modeled here using a periodic boundary condition on the left and right. 

\begin{table*}[ht!] 
\begin{center}
\caption{Stream parameters along with time of ignition. Half-width is $10^8$\ cm, density is $2.5 \times 10^4$\ g\ cm$^{-3}$, stream angle 35.5$^\circ$ and base density of the helium shell is $5 \times 10^5$\ g\ cm$^{-3}$. The peak density column represents the maximum density at the ignition point.\\
X - indicates no ignition. The non-igniting case was run to 14\ s without ignition. \\
** - High resolution run. cell size $\approx$ 10\ km \\
*** - High resolution run. cell size $\approx$ 5\ km  \\
*** - Burning limiter run.} 
\begin{tabular}{|c|c|c|c|c|}
\multicolumn{4}{c}{Pure helium shell - Magnesium core and small nuclear reaction network}                                                                   
\\ \hline
 \multicolumn{1}{|l|}{Velocity ($10^8$\ cm\ s$^{-1}$)} &  \multicolumn{1}{|l|}{Surface Length ($10^8$\ cm)} & \multicolumn{1}{|l|}{Time of impact (s)} & \multicolumn{1}{|l|}{Time of ignition (s)} & \multicolumn{1}{|l|}{Peak density ($10^6$\ g\ cm$^{-3}$)} \\ \hline 
 4.0                              & 30          & 1.3     &   X  & N/A                 \\ \hline
  5.0                              & 30         & 1.2      & 7.50  & 1.44               \\ \hline 
  5.0                            & 20          & 1.2     & 5.74 & 2.51                \\ \hline 
  5.5                              & 20         & 1.15      & 6.98  & 1.68               \\ \hline
  6.0                             & 30           & 1.13    & 6.92 & 1.43                 \\ \hline
  6.0                               & 20          & 1.13     & 4.30  & 2.09               \\ \hline
  6.0                               & 20          & 1.14     &  4.2* & 1.65                \\ \hline
  6.0                               & 20          & 1.2    &  4.67** & 1.89               \\ \hline
  7.0                               & 20           & 1.05    & 4.16  & 2.08               \\ \hline
  8.0                               & 20         & 1.0      & 3.61  & 1.30                   \\ \hline
  9.0                               & 20          & 0.95     & 3.30 & 1.49                \\ \hline
  10.0                               & 20          & 0.9     & 1.42 & 2.02                \\ \hline
 15.2                               & 20            & 0.7   & 1.00  & 2.85               \\ \hline
 17.2                               & 20             & 0.63   & 0.9 & 2.57                 \\ \hline

\end{tabular}
\newline

\begin{tabular}{|c|c|c|c|c|}
\multicolumn{4}{c}{Helium + Nitrogen shell - CO core and large nuclear reaction network}                                                                   
\\ \hline
 \multicolumn{1}{|l|}{Velocity ($10^8$\ cm\ s$^{-1}$)}  & \multicolumn{1}{|l|}{Surface Length ($10^8$\ cm)} & \multicolumn{1}{|l|}{Time of impact (s)} & \multicolumn{1}{|l|}{Time of ignition (s)} & \multicolumn{1}{|l|}{Peak density ($10^6$\ g\ cm$^{-3}$)}\\ \hline
 4.0                        & 20              & 1.3  & 6.05 & 0.71                    \\ \hline 
 5.5                       & 20               & 1.15 & 4.31  & 1.68  \\ \hline 
 6.0                        & 20              & 1.14  & 4.68   & 1.88                  \\ \hline
 6.0                        & 20              & 1.14  & 5.66***        & 0.98   \\ \hline
     
\end{tabular}
\newline
\label{table:2}
\end{center}
\end{table*}

The density plot at t=3.0\ s shows how the material, once wrapped around and coming in from the right, starts interacting with the right edge of the stream. The interaction of the wrapped around material with the stream introduces irregularity in the stream reminiscent of the stream behavior seen in the motivating work of \citet{Guillochonetal2010}. Over time the wrapped around material constricts the incoming material to a narrower stream thereby increasing the density of material impacting the surface. This leads to both a deeper penetration in the helium envelope and a compression of material from the incoming stream forming a high density region. Compression of material causes the region to heat up. When the temperature of this high density region reaches roughly $2 \times 10^9$\ K the surface is ignited and a detonation wave is produced. The propagating detonation can be seen in Figure \ref{fig:dens_approx13} where it is moving to the left at t=5.4\ s (position: x = $5 \times 10^8$\ cm, y = $2 \times 10^8$\ cm) and at t=5.7\ s (position: x = $2 \times 10^8$\ cm, y = $2 \times 10^8$\ cm). The temperature evolution near the time of ignition is shown in Figure \ref{fig:temp_approx13}. As seen in Table \ref{table:2}, the peak densities seem to mostly lie around $1.5 - 2 \times 10^6$\ g\ cm$^{-3}$ for the pure helium shell cases and around $1 - 2 \times 10^6$\ g\ cm$^{-3}$ for the helium with nitrogen cases. However, there is no trend in relation to the velocity or surface length which further demonstrates that ignition conditions rely on factors beyond just the peak density or temperature achieved on stream impact.
 
The deep penetration of the stream could dredge up material from the core. In such a scenario the core material will mix with the material in the shell. The presence of carbon and oxygen in the shell is known to detonate the surface with greater ease than in pure helium shells \citep{Shen_2014}. By design, this does not matter for the pure helium case since magnesium would be mixed into the pure helium shell and would not contribute to the occurrence of a detonation. However, the deeper penetration in the helium with nitrogen case could result in some of the core material (carbon and oxygen) to be dredged up and mixed with the outer shell. This is discussed in detail in Section \ref{sec:heshell}.

Two higher resolution runs were also performed for one of the pure helium shell cases as seen in Table \ref{table:2}. By doing so we were not attempting to find mesh convergence but were trying to investigate if these cases continue to ignite in an analogous manner at higher resolutions. As seen in Table \ref{table:2}, they consistently do so. In our simulations, the stream hits the surface of the WD and ignites a detonation. In the focusing mechanism cases, the wrapped around material interacts with the incoming stream to form ringlets (Kelvin-Helmholtz instability) in the stream (as seen in Figure \ref{fig:dens_approx13}). The impact of these local ringlet inhomogeneities in the stream appear to
lead to the ignition of the surface layer near the core-shell interface. With increased resolution, these ringlets become more distinct, clearer and more pronounced in structure (can be observed by comparing animations from the two different resolution cases in the Zenodo link at the end of the paper) but the fact that the ignition times are consistent shows that the process is not very sensitive to resolution. Since the development of the stream inhomogeneities is a random process driven by a flow instability, we expect some inherent stochasticity in the time of ignition, and deem the observed variation between resolutions to be within those expectations based on the typical sizes of the inhomogeneties and their velocities.
Although, other studies have investigated higher resolutions than a cell size of $\approx$ 5\ km \citep{Moore_2013, Rivas2022}, our goal here is not to do a resolution study but to present the results of our completed 2D study before we move on to 3D work.

In addition to the resolution study, one helium with nitrogen shell case was run with a burning limiter \citep{Kushnir_2013}, showing ignition as listed in Table \ref{table:2}.  While not determinative, this lends support to the robustness of the ignition. Our limiter is described in \citet{Boos_2021} and we use a value of ${|\Delta \ln \ T |}_{max}$ of 0.1, as done there (Refer to \citet{Boos_2021} for a detailed explanation of the burning limiter). The ignition takes place slightly later when the reactions are allowed to be restricted by the limiter, approximately when the next inhomogeneity in the stream interacts with the surface.  The site of the ignition in the absence of a limiter still shows the beginnings of a runaway, but the detonation is unable to propagate away from the ignition site, possibly due to the artificial thickening introduced by the limiter.  Since the underlying mechanism being discussed here is the continued focusing of the stream, we believe that a slightly later ignition is still a robust demonstration that such focusing can lead to ignition.


\begin{figure*}[ht!]
\centering
\includegraphics[width=\textwidth]{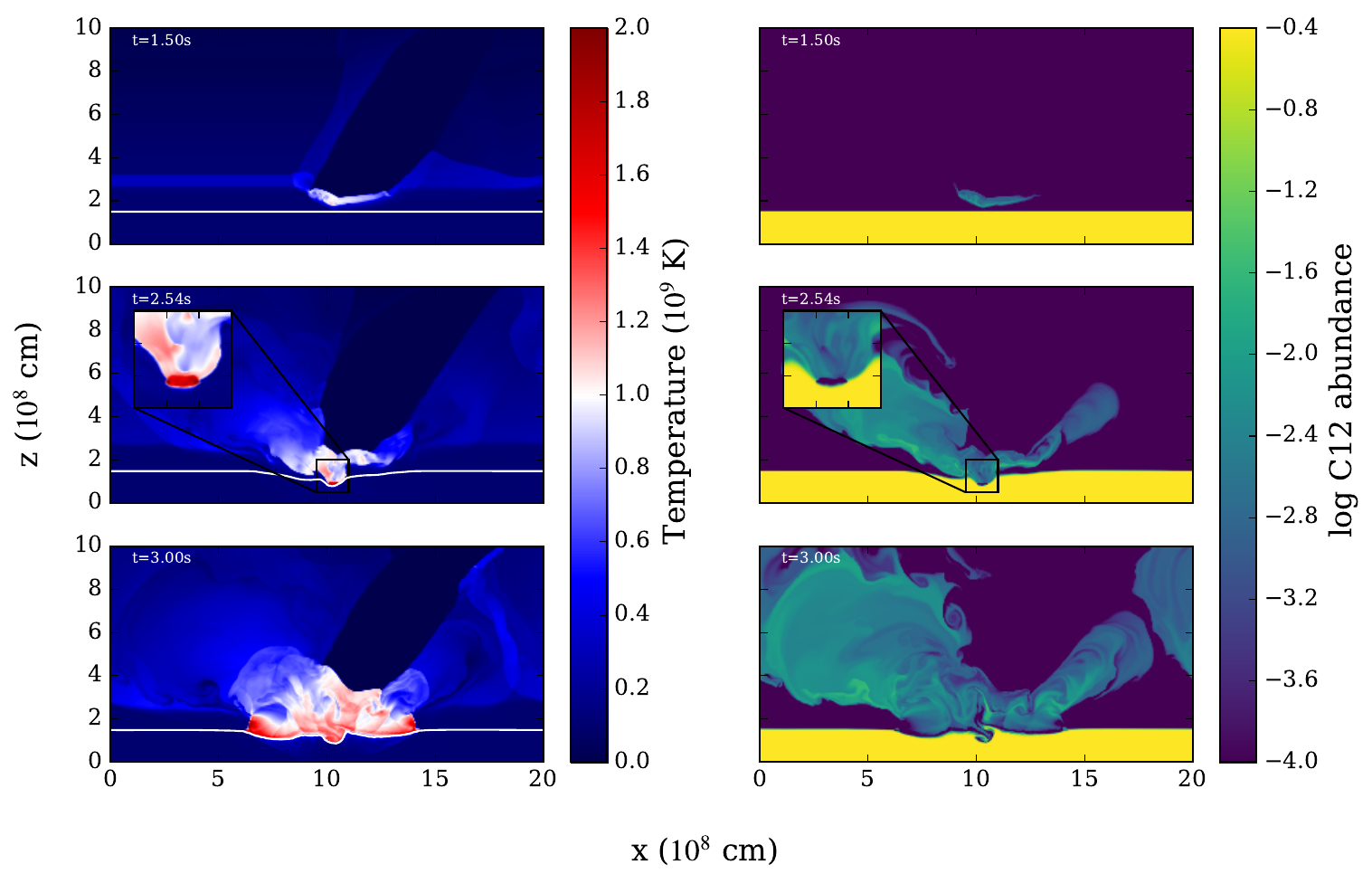}
\caption{Snapshots of temperature and carbon when the stream impacts the surface, at ignition and after ignition for a two-dimensional stream impacting the surface of a WD with a helium + nitrogen shell and a CO core. The density at the base of the helium layer is $3 \times 10^5$\ g\ cm$^{-3}$. At t = 1.50\ s the stream hits the surface of the WD. At t = 2.53\ s, the surface ignites. At t = 3.00\ s the detonation propagating on the surface can be observed. The white contour is plotted at a 0.35 $^{12}$C abundance and represents the interface between the core and shell of the WD.}
\label{fig:tempcomp_largenet}
\end{figure*}
 
Utilization of the larger nuclear network also contributes to an easier surface detonation. Data in Table \ref{table:2} shows that the velocity threshold was therefore lower than it was for the pure helium case when they both have a stream half-width of $1 \times 10^8$\ cm, an angle of 35.54$^\circ$ and a stream density of $2.5 \times 10^4$\ g\ cm$^{-3}$. The threshold velocity of the pure helium case was found to be $\approx 5 \times 10^8$\ cm\ s$^{-1}$ below which a surface detonation does not occur. The helium with nitrogen case detonates for a stream velocity as low as $4 \times 10^8$\ cm\ s$^{-1}$ and possibly for lower velocities. We did not attempt to find the velocity threshold because of the long computing time and our greater interest in varying the helium shell density.

\section{Varying the base density of the helium shell} \label{sec:heshell}

The double detonation of `thin' helium shell WDs in double degenerate systems produce light curves similar to normal SNIa \citep{Kromeretal, Townsley_2019, Boos_2021, Shen2021}. The elements formed in a SNIa determine what the maximum light spectrum would look like. But in a model such as ours, the elements formed would depend on the base helium shell density, thickness and mass. \citet{Townsley_2019} used a base helium shell density of $2 \times 10^5$\ g\ cm$^{-3}$, a shell mass of 0.021 $\textup{M}_\odot$ with $5\%$ $^{12}$C, $5\%$ $^{16}$O and $0.9\%$ $^{14}$N in the shell other than helium for the primary CO WD model which on double detonating produced spectra very similar to a normal SNIa at maximum light. Since a thin shell is required to produce a `normal' SNIa like spectra, it is necessary to check if our model can continue to detonate in the presence of a thin shell. 

We performed an additional study to determine if a stream impact ignition is possible for thin shell CO WDs too. Our models so far have used a base helium shell density of $5 \times 10^5$\ g\ cm$^{-3}$. Starting from this density, the base helium shell density is varied to lower values until the stream could not ignite the surface. Table \ref{table:3} lists the chosen base densities and the outcome. We ran all models until slightly after ignition since they were computationally demanding.

Shells with lower base densities ignite sooner than those with higher base densities.  A base helium shell density of $3 \times 10^5$\ g\ cm$^{-3}$ detonates at t=2.53\ s, compared to 4.68\ s for a base density of $5 \times 10^5$\ g\ cm$^{-3}$. In such a case, the ignition of the surface happens soon after the stream strikes the surface and the material from the stream has not wrapped around the star yet. A time sequence for this ``direct'' ignition case is shown in Figure \ref{fig:tempcomp_largenet}. Such an ignition takes place during the phase of evolution between the t=1.4\ s and t=3.7\ s panels shown in Figure \ref{fig:dens_approx13}. This is unlike our previous observation where the material from the stream wraps around the star and interacts with the incoming stream to eventually ignite the surface. 

 Thus, in the helium with nitrogen context, we have succeeded in realizing two distinct ignition modes. One involves interaction of the stream with previously deposited hot material, which leads to effective focusing of the stream thereby increasing the density at impact (referred to as the focusing mode). When the base shell density is decreased (thin helium layer) for the helium with nitrogen case, the surface ignition occurs promptly after the stream hits the surface (direct mode). Since the helium layer is thinner, the stream can penetrate the surface with a greater ease and compress the helium at the base of the helium layer and thereby ignite it. In the higher density cases we notice that the stream cannot penetrate the surface as easily, but the focusing of the stream helps in pushing the stream deeper and is therefore key for an ignition.

\begin{figure*}[ht!]
\centering
\includegraphics[width=\textwidth]{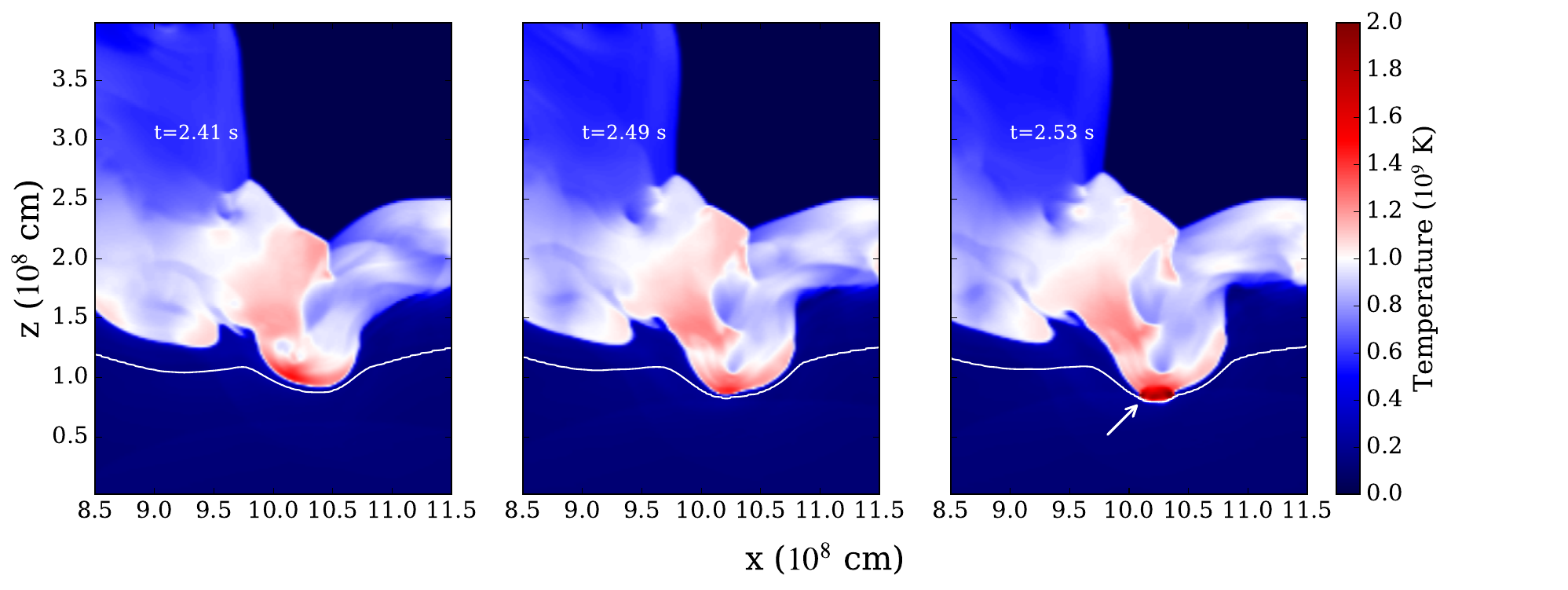}
\caption{Zoomed in temperature snapshots at the point of ignition of the helium with nitrogen case with $\rho_{\rm{base}}$ = $3 \times 10^5$\ g\ cm$^{-3}$. The three snapshots show the lead up to the ignition at the core-shell interface. The white contour is plotted at a 0.35 $^{12}$C abundance and represents the interface between the core and shell of the WD. The arrow points to the birth of a detonation. As seen in the plots, the incoming stream from the top compresses material at the interface to thereby ignite a detonation.}
\label{fig:ign_zoom}
\end{figure*}
 
 Zoomed in snapshots of the lead up to ignition are presented in Figure \ref{fig:ign_zoom} for the case with the base helium shell density of $3 \times 10^5$\ g\ cm$^{-3}$. As seen in the figure, the incoming stream compresses material at the core-shell interface which leads to the birth of a detonation. 

 The role of carbon dredge up or creation before detonation initiation is unclear. The carbon abundance plots in Figure \ref{fig:tempcomp_largenet} show both the products of pre-ignition helium burning and core material pushed out when the stream impacts the surface of the WD . At t = 1.5\ s, the stream has impacted the surface of the WD and is interacting with the helium shell. Therefore, the carbon abundance plot at this time suggests some production of carbon on impact ($\approx 10^{-2}$) due to helium burning. At t = 2.54\ s the stream has had time to penetrate into the underlying layers of the WD and as a result could suggest that the carbon seen on the left and right of the ignition region are from a combination of dredge up of core material and helium burning. Also, the absence of carbon at the point of ignition indicates carbon destruction by alpha capture and its contribution to igniting the surface.

We do not observe a clear threshold for the base helium shell density below which there is no ignition but instead observe shell densities that ignite but cannot propagate a detonation. A base helium shell density of $1.5 \times 10^5$\ g\ cm$^{-3}$ and $1.75 \times 10^5$\ g\ cm$^{-3}$ ignites the surface but fizzles out soon after. A base helium shell density of $1.9 \times 10^5$\ g\ cm$^{-3}$ ignites the surface and a detonation propagates to half the surface length. A base helium shell density of $3 \times 10^5$\ g\ cm$^{-3}$ ignites the surface and a detonation propagates around the entire WD. 
\citet{Shen_2014} have performed a detailed analysis of what shells might host a detonation, that is, use of a larger reaction network and using a more realistic envelope composition reduces the minimum envelope mass that can host a detonation. Therefore, we do not investigate further than finding that shell propagation is the important threshold.

\begin{figure*}[ht!]
\centering
\includegraphics[width=\textwidth]{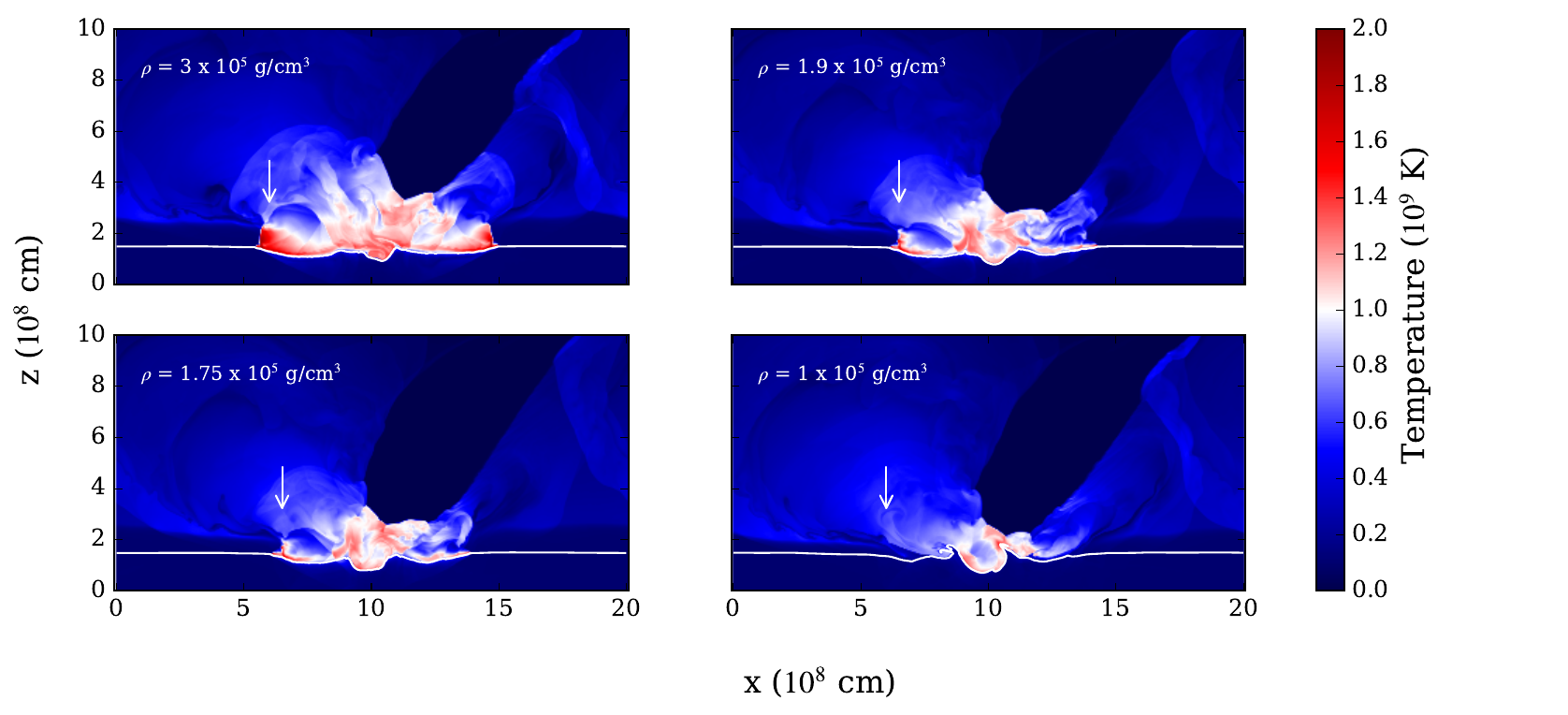}
\caption{Temperature snapshots of the helium with nitrogen case with different helium shell base densities approximately 3s after the simulation begins (the stream enters the domain). The base density of the model is indicated in the top left hand corner. From Table \ref{table:3} and from the annotated arrows that point to the shock front in} this plot, it is apparent that the strength of the detonation varies with the base densities. The model with $\rho_{\rm{base}}$ = $1.0 \times 10^5$\ g\ cm$^{-3}$ does not ignite while $\rho_{\rm{base}}$ = $1.75 \times 10^5$\ g\ cm$^{-3}$ and $\rho_{\rm{base}}$ = $1.9 \times 10^5$\ g\ cm$^{-3}$ cases ignite the surface, but the detonation propagates to varying distances before fizzling out. The model with $\rho_{\rm{base}}$ = $3.0 \times 10^5$\ g\ cm$^{-3}$ has a prominently strong shock front as compared to the other cases and continues to propagate over the entire domain. An animated plot of this evolution is available from the Zenodo link at the end of the manuscript.
\label{fig:diff_base_dens}
\end{figure*}

\begin{figure}[ht!]
\centering
\includegraphics[width=0.45\textwidth]{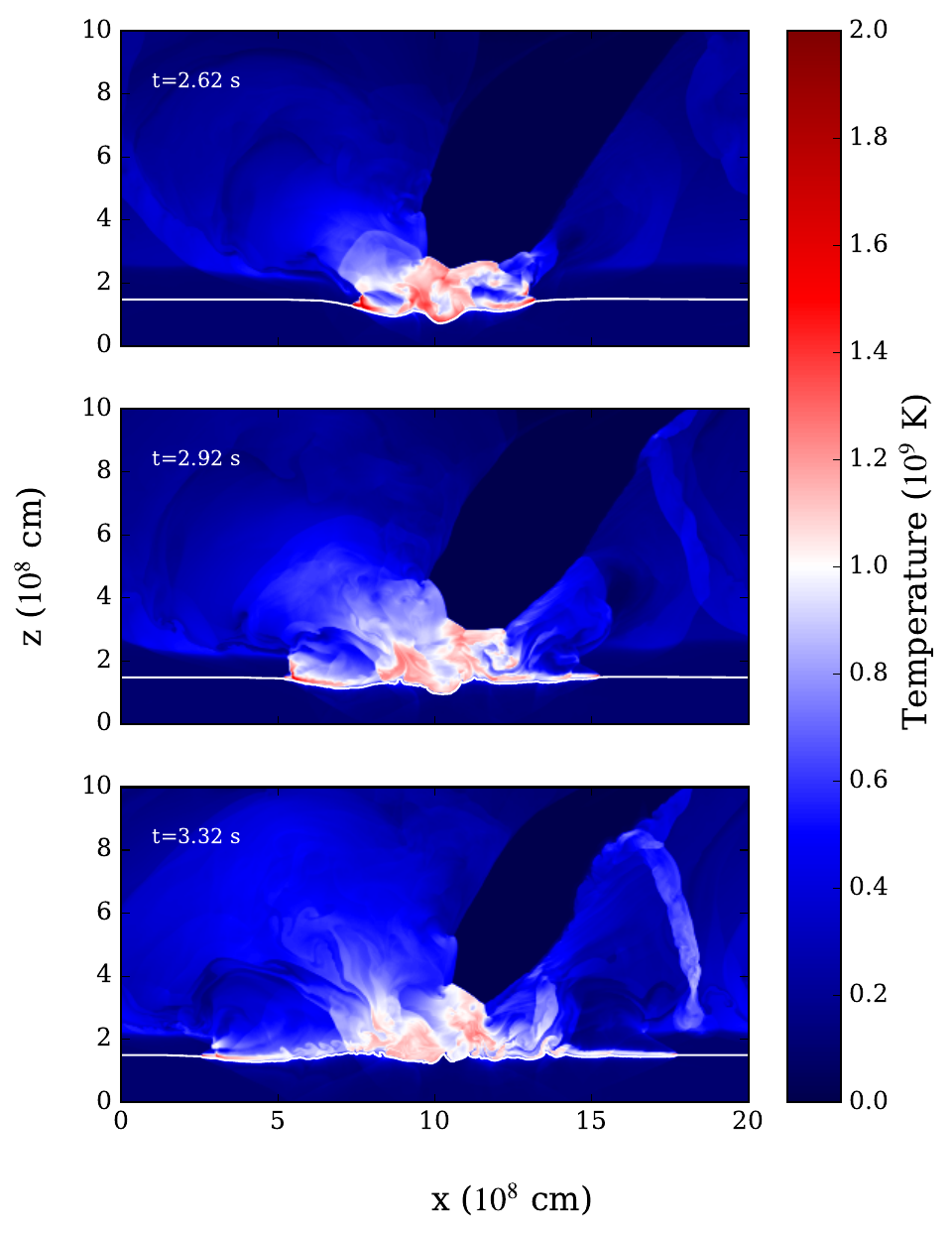}
\caption{Snapshots in temperature of the helium with nitrogen case with $\rho_{\rm{base}}$ = $1.9 \times 10^5$\ g\ cm$^{-3}$ model. These plots show the propagating detonation slowly fizzle out. The surface ignites by the direct mode and a detonation propagates, but the shock front separates from the burning front leading to the detonation fizzling out. The white contour is at a $^{12}$C abundance of 0.35 which also represents the core-shell interface. An animated plot of this evolution is available from the Zenodo link at the end of the manuscript.} 
\label{fig:propagate}
\end{figure}

 The peak densities at the ignition point and the ignition mode for each of the cases are also listed in Table \ref{table:3}. The peak density seems to slowly decrease with the base density of the helium shell and seems to lie between $1 - 2 \times 10^6$\ g\ cm$^{-3}$. A higher peak density may result in a stronger detonation compared to a lower peak density, which would also explain why the low helium shell base density is not able to sustain the detonation. The cases that ignite but are unable to sustain a detonation have weaker shockfronts compared to the cases that ignite and propagate a detonation. This can be seen in Figure \ref{fig:diff_base_dens} which compares four different cases from Table \ref{table:3} after ignition. Figure \ref{fig:propagate} demonstrates the dissipation of the detonation for one of the lower base helium shell density cases of $1.9 \times 10^5$\ g\ cm$^{-3}$. The shock front separates from the burning front which causes the detonation to fizzle out. Intermediate ignitions (ignition but no propagation) have also been observed in other work \citep{Iwata2022}.
 
  The detonation might also not be able to propagate on the shell due to how thin the helium layer is and because of the lack of carbon in the shell \citep{Moore_2013}. We also ran the case with a helium shell base density of $1.9 \times 10^5$\ g\ cm$^{-3}$ again after adding 5\% carbon and 5\% oxygen to the shell and observed that the detonation wave is stronger and is able to propagate all around the surface without fizzling out. The requirement to add some carbon and oxygen in order to maintain propagation at lower shell densities demonstrates the importance of these burning processes as outlined in detail in \citet{Shen_2014}. It also supports the inclusion of the modest fractions as was done in the full-star models of \citet{Townsley_2019} and \citet{Boos_2021}.
 
\begin{table*}
\centering
\caption{Helium + Nitrogen shell - CO core and large nuclear reaction network: Varying base density of the shell. Half-width of the stream is $10^8$\ cm, density of the stream is $2.5 \times 10^4$\ g\ cm$^{-3}$, Stream velocity is $6.0 \times 10^8$\ cm\ s$^{-1}$, Stream angle is 35.5$^\circ$, surface length is $20 \times 10^8$\ cm and $\rho_{base}$ is the density at the base of helium shell. The peak density column represents the maximum density at the point of ignition. \\
X - no ignition \\
** - 5\% carbon and 5\% oxygen was added to the shell. The detonation wave is stronger than in the case where no carbon and oxygen are present.\\
*** - detonation fizzles out before extending across the
entire domain}
    \begin{tabular}{|c|c|c|c|c|}

\hline

 \multicolumn{1}{|l|}{$\rho_{\rm{base}}$ ($10^5$\ g\ cm$^{-3}$)} & \multicolumn{1}{|l|}{Impact time (s)} & \multicolumn{1}{|l|}{Ignition time (s)}  & \multicolumn{1}{|l|}{Peak density ($10^6$\ g\ cm$^{-3}$)} & \multicolumn{1}{|l|}{Ignition mode}\\ \hline
        5      &  1.14 &    4.68 &  1.88 & focusing mechanism\\ \hline
        3      &  1.2 &    2.53  &  1.51 &  direct mechanism\\ \hline
        2      &  1.23 &    2.42** & 1.40 &  direct mechanism\\ \hline
        1.9    &  1.23 &    2.29** &  1.04 &  direct mechanism\\ \hline
        1.9*    &  1.23 &    2.30 &   0.88 &  direct mechanism\\ \hline
        1.75   &  1.23 &    2.33** &  0.98 &  direct mechanism   \\ \hline
        1.5    &  1.24 &    2.30** &   0.98 &  direct mechanism \\ \hline
        1      &  1.26 &    X  &   N/A &   N/A\\ \hline
  
 
\end{tabular}
\label{table:3}
\end{table*}

\section{Summary and Discussion} \label{sec:summary}

We present 2D hydrodynamic simulations of a stream of material, from the companion WD that was transferred to the primary WD on Roche-lobe overflow, impacting the surface of the primary WD and eventually producing a helium shell detonation. The helium shell detonation is the first step in the double-detonation model where the shell detonation triggers a secondary core detonation which blows up the entire star. Previous work \citep{Guillochonetal2010, Pakmor2013} has simulated this in 3D but the low resolution ($\approx$ 35\ km cell size in \citet{Guillochonetal2010} and $\approx$ 95\ km cell size everywhere in \citet{Pakmor2013} except $\approx$ 45\ km at the shell) made the ignition process unclear. Also, inclusion of the entire binary constrains choices of stream geometry
in ways that are undesirable for studying just the stream-surface interaction itself. Our work is focused on studying the ignition mechanism of the shell to further understand double-detonation models and the SN Ia ignition mechanism. 

Work presented here includes an initial study with a single circular mass impacting the surface to explain the behavior of an impact on the surface. A parameter study on impactor parameters like entry velocity, impactor radius and entry angle indicates the presence of a threshold in these parameters below which there is no surface ignition. The threshold itself varies depending on the chosen impactor density. This study was used to identify the stream parameters that would produce a surface detonation.

We then used these parameters in a 2D simulation where a stream of material impacts the surface. There were two types of setups that we used, the first being a `pure helium shell' setup with a small 13-nuclide alpha chain reaction network and a Mg core instead of a CO core WD, and the second being a `helium with nitrogen shell' with a larger reaction network (55-isotopes), a CO core and a helium shell and stream that contain the expected trace abundance of nitrogen. The first setup was an initial test to see if a surface detonation was possible and the second emulated a more realistic scenario. Both setups produce surface detonations for the initial parameters and for smaller impact velocities. In fact, the second setup with a more realistic system also produces a surface detonation when the shell density is decreased to  $1.9 \times 10^5$\ g\ cm$^{-3}$ (thin helium shell) further confirming the possibility of the double-detonation model in thin shell WDs. 

Two main mechanisms are observed. When the WD has a smaller helium shell base density, the surface detonates soon after the stream hits the surface (the direct mechanism). When the WD has a larger helium shell base density the detonation takes longer to occur and happens a while after the stream hits the surface (the focusing mechanism). In this case, the material from the stream that bounces up after impacting the WD slowly wraps around the surface and eventually interacts with the stream and compresses it. The stream is concentrated to a small width and therefore the density of material striking the surface is higher. These two mechanisms clearly indicate that (1) A surface detonation necessary for the double detonation mechanism is possible when mass transfer from the companion WD occurs (2) Thinner helium shells ignite more easily than thick ones because the stream can readily penetrate them, but the detonation propagates more effectively in thicker shells.  

Although we observe two different ignition modes, in actuality the surface ignition necessary for a double detonation to occur would probably be something in between these two modes. When mass transfer begins from the secondary WD, the flow of material to the primary WD gradually increases in strength unlike our simulation where the stream has a constant accretion rate. So in reality, the stream starts off weaker in intensity, directly impacts the primary WD, and wraps around the WD forming an envelope that continues to interact with incoming material. The incoming stream will grow in strength (increase in accretion rate) and the envelope formed from all the previous accretion will continue to interact with the stream which will lead to an ignition on the surface, similar to the focusing mechanism. Also, the helium ignition triggered in the thinner shell cases suggest that a detonation can only propagate in the presence of sufficient fuel. For the WD to double detonate, the detonation in the shell would have to be strong enough to trigger a secondary carbon detonation. For that reason, the shell cannot be too thin but also cannot be too thick because the burning of a thick helium shell will produce $^{56}$Ni on the outside of SNeIa ejecta \citep{Woosley1994, Hoeflich1996, Polin2019}. Other parameters like the stream density and stream velocity which were not varied for the thinner shell cases may also play a role in determining the exact ignition mechanism. The composition structure of the layers already on the WD appears to also be important \citep{Shen_2024}.

It is important to remember that these simulations were modelled in two dimensions. We have deferred characterizing the ignition process in detail because this process might be very different in 3D. The stream is currently a sheet since it is restricted to 2D and its impact on the surface of the WD also behaves like a sheet impacting the WD surface as opposed to a cylindrical stream. Also, the material that bounces off the surface, wraps around the star in just one direction because the simulation is two dimensional rather than moving in all directions as it would in 3D. This may change how the incoming stream is focused in the focusing mechanism. Also the formation of detonation shocks are likely to be different (harder) in 3D than in 2D. Therefore, follow up 3D studies are required to understand how these results extend to more realistic configurations.

Animated plots of the density and temperature evolution for all the helium with nitrogen shell models with varying shell base densities and one example case of the pure helium shell model are available in the following link. \url{https://zenodo.org/doi/10.5281/zenodo.13119353} 


\begin{acknowledgments}
We thank the referee for their constructive feedback, which greatly improved this paper. Early portions of this work were supported by the NASA Astrophysics Theory Program (NNX17AG28G).  Computations were performed at the University of Alabama's high performance computing center. D.M.T received support for this work from NASA grant HST-AR-16156. N.R. received support from the National Science Foundation under Grant No. 2307442. N.R also received support from NASA FINESST under award number 80NSSC23K1438. Financial support for K.J.S. was in part provided by NASA/ESA Hubble Space Telescope programs \#15871 and \#15918.
\end{acknowledgments}

%

\vspace{5mm}


\software{Flash 4.3 \citep{Fryxell, DUBEY2009512, Dubey2013, Dubey2014},  
          MESA \citep{Paxton2011, Paxton2013, Paxton2015, Paxton2018, Paxton2019, Jermyn2023}, 
          yt \citep{yt}
         }



\appendix

\section{Nuclear Timescale} \label{sec:tnuc}

\begin{figure}[ht!]
\centering
\includegraphics[width=\textwidth]{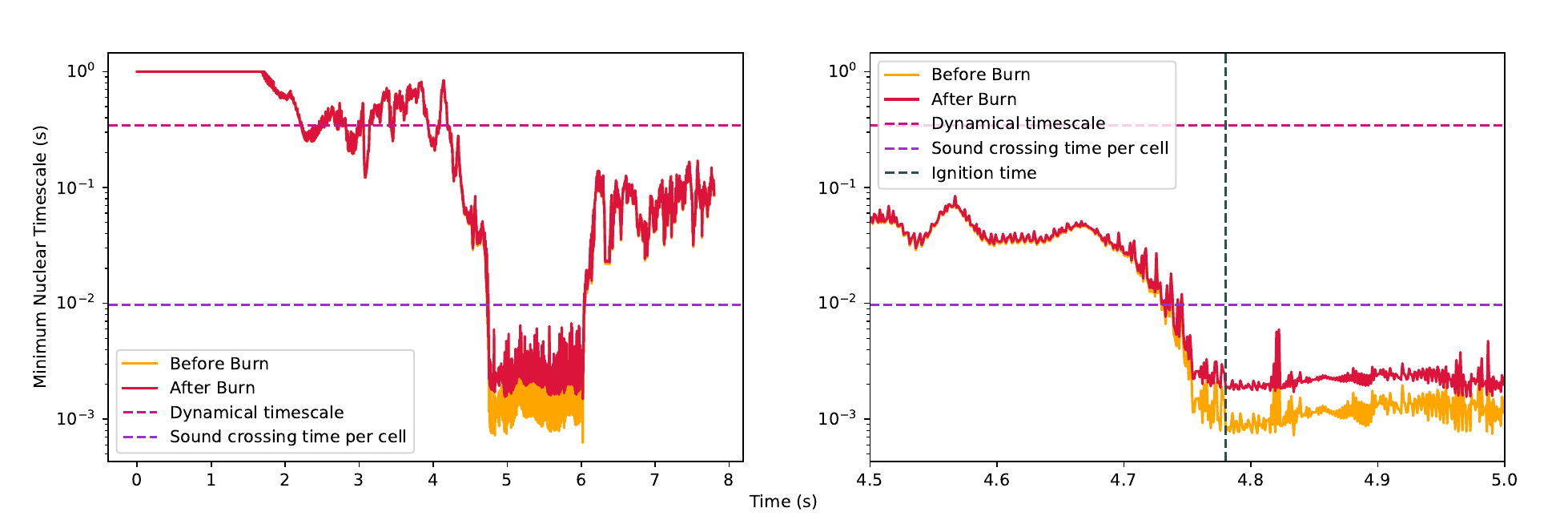}
\caption{Evolution of the minimum nuclear timescale in the domain over time. The plot on the right is a zoomed in version of the plot on the left to focus on the minimum nuclear timescale near ignition. The dynamical timescale and the sound crossing time of a cell are also shown in the plot as a pink and purple dashed line respectively. The minimum nuclear timescale seems to drop below the dynamical time scale before ignition ($\tau_{nuc}/\tau_{dyn} \ll 1$) and stays at an approximately constant value soon after the ignition as you would expect it to be for a propagating detonation. The detonation front crosses the whole domain by about 6\ s.} 
\label{fig:tnuc}
\end{figure}

When a detonation occurs the nuclear timescale is much smaller than the dynamical timescale \citep{Dan2012} or $\tau_{nuc}/\tau_{dyn} \ll 1$. $\tau_{nuc} = E_{in}/\dot E_{nuc}$ where $E_{in}$ is the internal energy and $\dot E_{nuc}$ is the nuclear energy generation rate and $t_{dyn} = H/ c_{s}$ where the scale height, $H\approx$ 700\ km and the speed of sound, $c_s \approx 2.05 \times 10^8$ \ cm\ s$^{-1}$ in our simulations. In Figure \ref{fig:tnuc} we plotted the minimum $\tau_{nuc}$ over time for one of the pure helium shell cases from Table \ref{table:2}, the case with velocity of the stream = $6.0 \times 10^8$\ cm\ s$^{-1}$, surface length = $20 \times 10^8$\ cm, density of stream = $2.5 \times 10^4$\ g\ cm$^{-3}$, half width of the stream = $1 \times 10^8$\ cm, stream angle = 35.5$^\circ$, base density of the helium shell = $5 \times 10^5$\ g\ cm$^{-3}$ and a cell size $\approx$ 20\ km resolution. For technical reasons, this was run with a newer version of our setup having
improved boundary conditions and a stream density that gradually decreases from the inner to the outer part of the stream. This results in a similar but not precisely identical ignition, due to minor numerical differences.
The plot on the left is the evolution of the minimum $\tau_{nuc}$ over time while the one on the right is the evolution near the point of ignition (at around 4.78\ s ). Flash is operator split such that energy deposition and hydrodynamic flow are computed sequentially. This results in $E_{in}$ being different just before the energy deposition rate is computed and just before hydrodynamic flow is computed (after energy deposition). Two curves are plotted, one where $\tau_{nuc}$ was calculated before the nuclear burning and one calculated after the nuclear burning when energy is deposited to $E_{in}$ from the burning. We can see in this figure how the minimum $\tau_{nuc}$ drastically decreases below $\tau_{dyn}$ before ignition and flattens out to a consistent value  after ignition as expected for a propagating detonation. The minimum $\tau_{nuc}$ is also lesser than the sound crossing time per cell which is expected since our simulations are still under-resolved. The detonation propagation crosses the entire domain by about 6\ s, so that there is no longer an active detonation front after that, and $\tau_{nuc}$ rises back up. Ongoing burning continues in the hot material, but without the
detonation front.

\bibliography{ref}{}

\begin{thebibliography}{}
\expandafter\ifx\csname natexlab\endcsname\relax\def\natexlab#1{#1}\fi
\providecommand{\url}[1]{\href{#1}{#1}}
\providecommand{\dodoi}[1]{doi:~\href{http://doi.org/#1}{\nolinkurl{#1}}}
\providecommand{\doeprint}[1]{\href{http://ascl.net/#1}{\nolinkurl{http://ascl.net/#1}}}
\providecommand{\doarXiv}[1]{\href{https://arxiv.org/abs/#1}{\nolinkurl{https://arxiv.org/abs/#1}}}

\bibitem[{Boos {et~al.}(2021)Boos, Townsley, Shen, Caldwell, \& Miles}]{Boos_2021}
Boos, S.~J., Townsley, D.~M., Shen, K.~J., Caldwell, S., \& Miles, B.~J. 2021, The Astrophysical Journal, 919, 126, \dodoi{10.3847/1538-4357/ac07a2}

\bibitem[{{Collins} {et~al.}(2022){Collins}, {Gronow}, {Sim}, \& {R{\"o}pke}}]{Collins2022}
{Collins}, C.~E., {Gronow}, S., {Sim}, S.~A., \& {R{\"o}pke}, F.~K. 2022, \mnras, 517, 5289, \dodoi{10.1093/mnras/stac2665}

\bibitem[{Cortes \& Vapnik(1995)}]{SVC}
Cortes, C., \& Vapnik, V. 1995, Machine learning, 20, 273

\bibitem[{{Dan} {et~al.}(2015){Dan}, {Guillochon}, {Br{\"u}ggen}, {Ramirez-Ruiz}, \& {Rosswog}}]{Dan2015}
{Dan}, M., {Guillochon}, J., {Br{\"u}ggen}, M., {Ramirez-Ruiz}, E., \& {Rosswog}, S. 2015, \mnras, 454, 4411, \dodoi{10.1093/mnras/stv2289}

\bibitem[{{Dan} {et~al.}(2011){Dan}, {Rosswog}, {Guillochon}, \& {Ramirez-Ruiz}}]{Dan2011}
{Dan}, M., {Rosswog}, S., {Guillochon}, J., \& {Ramirez-Ruiz}, E. 2011, \apj, 737, 89, \dodoi{10.1088/0004-637X/737/2/89}

\bibitem[{Dan {et~al.}(2012)Dan, Rosswog, Guillochon, \& Ramirez-Ruiz}]{Dan2012}
Dan, M., Rosswog, S., Guillochon, J., \& Ramirez-Ruiz, E. 2012, Monthly Notices of the Royal Astronomical Society, 422, 2417, \dodoi{10.1111/j.1365-2966.2012.20794.x}

\bibitem[{Dubey {et~al.}(2009)Dubey, Antypas, Ganapathy, Reid, Riley, Sheeler, Siegel, \& Weide}]{DUBEY2009512}
Dubey, A., Antypas, K., Ganapathy, M.~K., {et~al.} 2009, Parallel Computing, 35, 512, \dodoi{https://doi.org/10.1016/j.parco.2009.08.001}

\bibitem[{Dubey {et~al.}(2013)Dubey, Calder, Daley, Fisher, Graziani, Jordan, Lamb, Reid, Townsley, \& Weide}]{Dubey2013}
Dubey, A., Calder, A.~C., Daley, C., {et~al.} 2013, The International Journal of High Performance Computing Applications, 27, 360, \dodoi{10.1177/1094342012464404}

\bibitem[{Dubey {et~al.}(2014)Dubey, Antypas, Calder, Daley, Fryxell, Gallagher, Lamb, Lee, Olson, Reid, Rich, Ricker, Riley, Rosner, Siegel, Taylor, Weide, Timmes, Vladimirova, \& ZuHone}]{Dubey2014}
Dubey, A., Antypas, K., Calder, A.~C., {et~al.} 2014, The International Journal of High Performance Computing Applications, 28, 225, \dodoi{10.1177/1094342013505656}

\bibitem[{{Fenn} {et~al.}(2016){Fenn}, {Plewa}, \& {Gawryszczak}}]{Fenn2016}
{Fenn}, D., {Plewa}, T., \& {Gawryszczak}, A. 2016, \mnras, 462, 2486, \dodoi{10.1093/mnras/stw1831}

\bibitem[{{Fryxell} {et~al.}(2000){Fryxell}, {Olson}, {Ricker}, {Timmes}, {Zingale}, {Lamb}, {MacNeice}, {Rosner}, {Truran}, \& {Tufo}}]{Fryxell}
{Fryxell}, B., {Olson}, K., {Ricker}, P., {et~al.} 2000, \apjs, 131, 273, \dodoi{10.1086/317361}

\bibitem[{García-Senz {et~al.}(2018)García-Senz, Cabezón, \& Domínguez}]{Garcia-Senz_2018}
García-Senz, D., Cabezón, R.~M., \& Domínguez, I. 2018, The Astrophysical Journal, 862, 27, \dodoi{10.3847/1538-4357/aacb7d}

\bibitem[{{Glasner} {et~al.}(2018){Glasner}, {Livne}, {Steinberg}, {Yalinewich}, \& {Truran}}]{Glasner2018}
{Glasner}, S.~A., {Livne}, E., {Steinberg}, E., {Yalinewich}, A., \& {Truran}, J.~W. 2018, \mnras, 476, 2238, \dodoi{10.1093/mnras/sty421}

\bibitem[{{Gronow} {et~al.}(2020){Gronow}, {Collins}, {Ohlmann}, {Pakmor}, {Kromer}, {Seitenzahl}, {Sim}, \& {R{\"o}pke}}]{Gronow2020}
{Gronow}, S., {Collins}, C., {Ohlmann}, S.~T., {et~al.} 2020, \aap, 635, A169, \dodoi{10.1051/0004-6361/201936494}

\bibitem[{Gronow {et~al.}(2021)Gronow, Collins, Sim, \& Ropke}]{Gronowetal2021}
Gronow, S., Collins, C.~E., Sim, S.~A., \& Ropke, F.~K. 2021, A\&A, 649, A155, \dodoi{10.1051/0004-6361/202039954}

\bibitem[{Guillochon {et~al.}(2010)Guillochon, Ramirez-Ruiz, Dan, \& Rosswog}]{Guillochonetal2010}
Guillochon, J., Ramirez-Ruiz, E., Dan, M., \& Rosswog, S. 2010, Astrophysical Journal Letters, 709, \dodoi{10.1088/2041-8205/709/1/L64}

\bibitem[{{Hamada} \& {Salpeter}(1961)}]{Hamada1961}
{Hamada}, T., \& {Salpeter}, E.~E. 1961, \apj, 134, 683, \dodoi{10.1086/147195}

\bibitem[{{Hoeflich} \& {Khokhlov}(1996)}]{Hoeflich1996}
{Hoeflich}, P., \& {Khokhlov}, A. 1996, \apj, 457, 500, \dodoi{10.1086/176748}

\bibitem[{{Holcomb} {et~al.}(2013){Holcomb}, {Guillochon}, {De Colle}, \& {Ramirez-Ruiz}}]{Holcombe2013}
{Holcomb}, C., {Guillochon}, J., {De Colle}, F., \& {Ramirez-Ruiz}, E. 2013, \apj, 771, 14, \dodoi{10.1088/0004-637X/771/1/14}

\bibitem[{{Iwata} \& {Maeda}(2022)}]{Iwata2022}
{Iwata}, K., \& {Maeda}, K. 2022, \apj, 941, 87, \dodoi{10.3847/1538-4357/aca013}

\bibitem[{{Jermyn} {et~al.}(2023){Jermyn}, {Bauer}, {Schwab}, {Farmer}, {Ball}, {Bellinger}, {Dotter}, {Joyce}, {Marchant}, {Mombarg}, {Wolf}, {Sunny Wong}, {Cinquegrana}, {Farrell}, {Smolec}, {Thoul}, {Cantiello}, {Herwig}, {Toloza}, {Bildsten}, {Townsend}, \& {Timmes}}]{Jermyn2023}
{Jermyn}, A.~S., {Bauer}, E.~B., {Schwab}, J., {et~al.} 2023, \apjs, 265, 15, \dodoi{10.3847/1538-4365/acae8d}

\bibitem[{{Kromer} {et~al.}(2010){Kromer}, {Sim}, {Fink}, {R{\"o}pke}, {Seitenzahl}, \& {Hillebrandt}}]{Kromeretal}
{Kromer}, M., {Sim}, S.~A., {Fink}, M., {et~al.} 2010, \apj, 719, 1067, \dodoi{10.1088/0004-637X/719/2/1067}

\bibitem[{{Kushnir} {et~al.}(2013){Kushnir}, {Katz}, {Dong}, {Livne}, \& {Fern{\'a}ndez}}]{Kushnir_2013}
{Kushnir}, D., {Katz}, B., {Dong}, S., {Livne}, E., \& {Fern{\'a}ndez}, R. 2013, \apjl, 778, L37, \dodoi{10.1088/2041-8205/778/2/L37}

\bibitem[{{Lubow} \& {Shu}(2014)}]{Lubow1975}
{Lubow}, S.~H., \& {Shu}, F.~H. 2014, \apj, 788, 95, \dodoi{10.1088/0004-637X/788/1/95}

\bibitem[{{Maoz} {et~al.}(2014){Maoz}, {Mannucci}, \& {Nelemans}}]{Maoz}
{Maoz}, D., {Mannucci}, F., \& {Nelemans}, G. 2014, Annual Review of Astronomy and Astrophysics, 52, 107, \dodoi{10.1146/annurev-astro-082812-141031}

\bibitem[{Marsh {et~al.}(2004)Marsh, Nelemans, \& Steeghs}]{Marsh_2004}
Marsh, T.~R., Nelemans, G., \& Steeghs, D. 2004, Monthly Notices of the Royal Astronomical Society, 350, 113, \dodoi{10.1111/j.1365-2966.2004.07564.x}

\bibitem[{{Moll} \& {Woosley}(2013)}]{MollWoosley2013}
{Moll}, R., \& {Woosley}, S.~E. 2013, \apj, 774, 137, \dodoi{10.1088/0004-637X/774/2/137}

\bibitem[{Moore {et~al.}(2013)Moore, Townsley, \& Bildsten}]{Moore_2013}
Moore, K., Townsley, D.~M., \& Bildsten, L. 2013, The Astrophysical Journal, 776, 97, \dodoi{10.1088/0004-637X/776/2/97}

\bibitem[{{Nelemans} {et~al.}(2001){Nelemans}, {Portegies Zwart}, {Verbunt}, \& {Yungelson}}]{Nelemans2001}
{Nelemans}, G., {Portegies Zwart}, S.~F., {Verbunt}, F., \& {Yungelson}, L.~R. 2001, \aap, 368, 939, \dodoi{10.1051/0004-6361:20010049}

\bibitem[{{Paczy{\'n}ski}(1967)}]{1967AcA....17..287P}
{Paczy{\'n}ski}, B. 1967, \actaa, 17, 287

\bibitem[{{Pakmor} {et~al.}(2013){Pakmor}, {Kromer}, {Taubenberger}, \& {Springel}}]{Pakmor2013}
{Pakmor}, R., {Kromer}, M., {Taubenberger}, S., \& {Springel}, V. 2013, \apjl, 770, L8, \dodoi{10.1088/2041-8205/770/1/L8}

\bibitem[{{Pakmor} {et~al.}(2021){Pakmor}, {Zenati}, {Perets}, \& {Toonen}}]{Pakmor2021}
{Pakmor}, R., {Zenati}, Y., {Perets}, H.~B., \& {Toonen}, S. 2021, \mnras, 503, 4734, \dodoi{10.1093/mnras/stab686}

\bibitem[{{Pakmor} {et~al.}(2022){Pakmor}, {Callan}, {Collins}, {de Mink}, {Holas}, {Kerzendorf}, {Kromer}, {Neunteufel}, {O'Brien}, {R{\"o}pke}, {Ruiter}, {Seitenzahl}, {Shingles}, {Sim}, \& {Taubenberger}}]{Pakmor2022}
{Pakmor}, R., {Callan}, F.~P., {Collins}, C.~E., {et~al.} 2022, \mnras, 517, 5260, \dodoi{10.1093/mnras/stac3107}

\bibitem[{{Paxton} {et~al.}(2011){Paxton}, {Bildsten}, {Dotter}, {Herwig}, {Lesaffre}, \& {Timmes}}]{Paxton2011}
{Paxton}, B., {Bildsten}, L., {Dotter}, A., {et~al.} 2011, \apjs, 192, 3, \dodoi{10.1088/0067-0049/192/1/3}

\bibitem[{{Paxton} {et~al.}(2013){Paxton}, {Cantiello}, {Arras}, {Bildsten}, {Brown}, {Dotter}, {Mankovich}, {Montgomery}, {Stello}, {Timmes}, \& {Townsend}}]{Paxton2013}
{Paxton}, B., {Cantiello}, M., {Arras}, P., {et~al.} 2013, \apjs, 208, 4, \dodoi{10.1088/0067-0049/208/1/4}

\bibitem[{{Paxton} {et~al.}(2015){Paxton}, {Marchant}, {Schwab}, {Bauer}, {Bildsten}, {Cantiello}, {Dessart}, {Farmer}, {Hu}, {Langer}, {Townsend}, {Townsley}, \& {Timmes}}]{Paxton2015}
{Paxton}, B., {Marchant}, P., {Schwab}, J., {et~al.} 2015, \apjs, 220, 15, \dodoi{10.1088/0067-0049/220/1/15}

\bibitem[{{Paxton} {et~al.}(2018){Paxton}, {Schwab}, {Bauer}, {Bildsten}, {Blinnikov}, {Duffell}, {Farmer}, {Goldberg}, {Marchant}, {Sorokina}, {Thoul}, {Townsend}, \& {Timmes}}]{Paxton2018}
{Paxton}, B., {Schwab}, J., {Bauer}, E.~B., {et~al.} 2018, \apjs, 234, 34, \dodoi{10.3847/1538-4365/aaa5a8}

\bibitem[{{Paxton} {et~al.}(2019){Paxton}, {Smolec}, {Schwab}, {Gautschy}, {Bildsten}, {Cantiello}, {Dotter}, {Farmer}, {Goldberg}, {Jermyn}, {Kanbur}, {Marchant}, {Thoul}, {Townsend}, {Wolf}, {Zhang}, \& {Timmes}}]{Paxton2019}
{Paxton}, B., {Smolec}, R., {Schwab}, J., {et~al.} 2019, \apjs, 243, 10, \dodoi{10.3847/1538-4365/ab2241}

\bibitem[{{Polin} {et~al.}(2019){Polin}, {Nugent}, \& {Kasen}}]{Polin2019}
{Polin}, A., {Nugent}, P., \& {Kasen}, D. 2019, \apj, 873, 84, \dodoi{10.3847/1538-4357/aafb6a}

\bibitem[{{Raskin} {et~al.}(2012){Raskin}, {Scannapieco}, {Fryer}, {Rockefeller}, \& {Timmes}}]{Raskin2012}
{Raskin}, C., {Scannapieco}, E., {Fryer}, C., {Rockefeller}, G., \& {Timmes}, F.~X. 2012, \apj, 746, 62, \dodoi{10.1088/0004-637X/746/1/62}

\bibitem[{{Rivas} {et~al.}(2022){Rivas}, {Harris}, {Hix}, \& {Messer}}]{Rivas2022}
{Rivas}, F., {Harris}, J.~A., {Hix}, W.~R., \& {Messer}, O.~E.~B. 2022, \apj, 937, 2, \dodoi{10.3847/1538-4357/ac8b06}

\bibitem[{{Roy} {et~al.}(2022){Roy}, {Tiwari}, {Bobrick}, {Kosakowski}, {Fisher}, {Perets}, {Kashyap}, {Lor{\'e}n-Aguilar}, \& {Garc{\'\i}a-Berro}}]{Roy2022}
{Roy}, N.~C., {Tiwari}, V., {Bobrick}, A., {et~al.} 2022, \apjl, 932, L24, \dodoi{10.3847/2041-8213/ac75e7}

\bibitem[{{Shen} {et~al.}(2024){Shen}, {Boos}, \& {Townsley}}]{Shen_2024}
{Shen}, K.~J., {Boos}, S.~J., \& {Townsley}, D.~M. 2024, arXiv e-prints, arXiv:2405.19417, \dodoi{10.48550/arXiv.2405.19417}

\bibitem[{{Shen} {et~al.}(2021){Shen}, {Boos}, {Townsley}, \& {Kasen}}]{Shen2021}
{Shen}, K.~J., {Boos}, S.~J., {Townsley}, D.~M., \& {Kasen}, D. 2021, \apj, 922, 68, \dodoi{10.3847/1538-4357/ac2304}

\bibitem[{Shen \& Moore(2014)}]{Shen_2014}
Shen, K.~J., \& Moore, K. 2014, The Astrophysical Journal, 797, 46, \dodoi{10.1088/0004-637X/797/1/46}

\bibitem[{{Shen} {et~al.}(2018){Shen}, {Boubert}, {G{\"a}nsicke}, {Jha}, {Andrews}, {Chomiuk}, {Foley}, {Fraser}, {Gromadzki}, {Guillochon}, {Kotze}, {Maguire}, {Siebert}, {Smith}, {Strader}, {Badenes}, {Kerzendorf}, {Koester}, {Kromer}, {Miles}, {Pakmor}, {Schwab}, {Toloza}, {Toonen}, {Townsley}, \& {Williams}}]{Shen2018b}
{Shen}, K.~J., {Boubert}, D., {G{\"a}nsicke}, B.~T., {et~al.} 2018, \apj, 865, 15, \dodoi{10.3847/1538-4357/aad55b}

\bibitem[{{Tanikawa} {et~al.}(2019){Tanikawa}, {Nomoto}, {Nakasato}, \& {Maeda}}]{Tanikawa2019}
{Tanikawa}, A., {Nomoto}, K., {Nakasato}, N., \& {Maeda}, K. 2019, \apj, 885, 103, \dodoi{10.3847/1538-4357/ab46b6}

\bibitem[{{Timmes} \& {Niemeyer}(2000)}]{Timmes2000}
{Timmes}, F.~X., \& {Niemeyer}, J.~C. 2000, \apj, 537, 993, \dodoi{10.1086/309043}

\bibitem[{Townsley {et~al.}(2019)Townsley, Miles, Shen, \& Kasen}]{Townsley_2019}
Townsley, D.~M., Miles, B.~J., Shen, K.~J., \& Kasen, D. 2019, The Astrophysical Journal Letters, 878, L38, \dodoi{10.3847/2041-8213/ab27cd}

\bibitem[{{Townsley} {et~al.}(2012){Townsley}, {Moore}, \& {Bildsten}}]{Townsley2012}
{Townsley}, D.~M., {Moore}, K., \& {Bildsten}, L. 2012, \apj, 755, 4, \dodoi{10.1088/0004-637X/755/1/4}

\bibitem[{{Turk} {et~al.}(2011){Turk}, {Smith}, {Oishi}, {Skory}, {Skillman}, {Abel}, \& {Norman}}]{yt}
{Turk}, M.~J., {Smith}, B.~D., {Oishi}, J.~S., {et~al.} 2011, The Astrophysical Journal Supplement Series, 192, 9, \dodoi{10.1088/0067-0049/192/1/9}

\bibitem[{{Wong} \& {Bildsten}(2023)}]{Wong2023}
{Wong}, T. L.~S., \& {Bildsten}, L. 2023, \apj, 951, 28, \dodoi{10.3847/1538-4357/acce9d}

\bibitem[{{Woosley} \& {Weaver}(1986)}]{WoosleyWeaver}
{Woosley}, S.~E., \& {Weaver}, T.~A. 1986, \araa, 24, 205, \dodoi{10.1146/annurev.aa.24.090186.001225}

\bibitem[{{Woosley} \& {Weaver}(1994)}]{Woosley1994}
---. 1994, \apj, 423, 371, \dodoi{10.1086/173813}

\bibitem[{{Zingale} {et~al.}(2002){Zingale}, {Dursi}, {ZuHone}, {Calder}, {Fryxell}, {Plewa}, {Truran}, {Caceres}, {Olson}, {Ricker}, {Riley}, {Rosner}, {Siegel}, {Timmes}, \& {Vladimirova}}]{Zingale2002}
{Zingale}, M., {Dursi}, L.~J., {ZuHone}, J., {et~al.} 2002, \apjs, 143, 539, \dodoi{10.1086/342754}

\end{thebibliography}
\bibliographystyle{aasjournal}


\end{document}